\pdfoutput=1
\documentclass[reprint,amsmath,amssymb,prd,nofootinbib,superscriptaddress]{revtex4-1}

\usepackage{aas_macros}
\usepackage{physics}
\usepackage{graphicx}
\usepackage{dcolumn}
\usepackage{fontawesome}
\usepackage{bm}
\usepackage[normalem]{ulem}

\usepackage{xcolor}

\definecolor{colorLink}{rgb}{0.9,0,0} % red
\definecolor{colorCite}{rgb}{0,0.7,0} % green
\definecolor{colorURL} {rgb}{0,0,0.8} % navy
\usepackage[colorlinks=true,linktocpage=true,linkcolor=colorLink,citecolor=colorCite,urlcolor=colorURL]{hyperref}
\newcommand{\HI}{H\textsc{i}\,\,}
\newcommand{\HII}{H\textsc{i}}
\defcitealias{WadFar21}{WF21}
\defcitealias{Kim20}{K21} 
\newcommand{\be}{\begin{equation}}
\newcommand{\ee}{\end{equation}}
\newcommand{\e}{$e^\pm$\ }
\newcommand{\tc}{$t_\mathrm{confine}$}
\def\WF{\citetalias{WadFar21}}
\def\Kim{\citetalias{Kim20}}

\newcommand{\lagr}{\mathcal{L}}
\newcommand{\DM}{\mathrm{DM}}
\newcommand{\taumin}{\tau^{\mathrm{min}}}

\newcommand{\alf}{Alfv$\acute{\text{e}}$n\ }

\begin{document}

\title{Strong constraints on decay and annihilation of dark matter\\
from heating of gas-rich dwarf galaxies}
\author{Digvijay Wadekar}
 \email{jayw@ias.edu}
\affiliation{Center for Cosmology and Particle Physics, Department of Physics, New York University, New York, NY 10003, USA}
\affiliation{School of Natural Sciences, Institute for Advanced Study, Princeton, NJ 08540, USA}
\author{Zihui Wang}
%\email{zihui.wang@nyu.edu}
\affiliation{Center for Cosmology and Particle Physics, Department of Physics, New York University, New York, NY 10003, USA}
\date{November 11, 2021}

\begin{abstract}
Gas-rich dwarf galaxies located outside the virial radius of their host are relatively pristine systems and have ultra-low gas cooling rates. This makes them very sensitive to heat injection by annihilation or decay of dark matter (DM). Such dwarfs are particularly sensitive to DM producing e$^\pm$ with energies 1$-$100 MeV or photons with energies 13.6 eV$-1$ keV, because these products can be efficiently thermalized in the neutral hydrogen gas of the dwarfs.
Following the methodology of \citet{WadFar21}, we require the rate of heat injection by DM to not exceed the ultra-low radiative cooling rate of gas in the Leo~T dwarf galaxy.
This gives model-independent bounds on $(i)$ the decay lifetime of DM to $e^\pm$ (photons) which are stronger than all the previous literature
for $m_\textup{DM}\sim$ 1$-$10 MeV ($m_\textup{DM}\sim0.02-$1 keV), $(ii)$ annihilation of DM to $e^{\pm}$ comparable to constraints from CMB and X/$\gamma$-ray surveys. We also translate our bounds for the case of the following DM models: axion-like particles (ALPs), sterile neutrinos, excited DM states, higgs portal scalars and dark baryons. 
Observations of gas-rich low-mass dwarfs from upcoming 21cm and optical surveys can therefore be powerful probes of a multitude of models of DM.  \href{https://github.com/JayWadekar/Gas_rich_dwarfs}{\faGithub}

\end{abstract}
\maketitle

\section{Introduction}

Detection of dwarf galaxies in our local group has accelerated at a tremendous pace with the advent of digital surveys like SDSS and DES (see Fig.~1 of \cite{Sim19}), and the trend is expected to continue with upcoming surveys like Rubin observatory \cite{LSST,Drl19_LSST,Mut21}, Roman telescope \cite{WFIRST} and Hyper Suprime Cam (HSC) \cite{Aih18} survey. Low mass dwarf galaxies are pristine laboratories to study effects of dark matter due to the baryonic feedback in them being weak, and the ratio of DM to baryonic mass in them being large. Furthermore, many popular alternatives to the cold dark matter model can affect the properties of these dwarfs (e.g.,~\cite{Dal22,Zav13,Jia21,Nad21,Nad19}).
% One long standing motivation to search for dwarf galaxies has been due to the ``missing satellites" problem \cite{BucPet18}.

The presence of neutral hydrogen (\HII) gas inside low mass dwarfs is related to their position with respect to their host: gas-less dwarfs are typically observed within the virial radius of the halo of their host, and vice versa for gas-rich dwarfs. This spatial dependence arises because gas in the dwarfs gets stripped due to pressure of the ionized medium of their host once they fall inside its virial radius (i.e., the ram pressure stripping effect \cite{GunGot72,Grc09, Spe14, Put21}). Observations of gas-rich dwarfs has been difficult in the past owing to their low luminosities because of being located beyond the Milky Way virial radius. However, recent high-resolution optical and 21cm surveys have discovered and characterized a large number of these dwarfs \cite{Wal08_Things,BegChe08_Figgs,Can11_Sheild,Ott12_Angst,Hun12_LittleThings,Ada13,Sau12,Sim19}. 

Finding gas-rich low-mass dwarfs is useful for testing galaxy scaling relations like the baryonic Tully-Fisher relation (BTFR \cite{McG00,Man19,McQ20,Man20}) or the mass metallicity relation at the very low-mass end.
% In the past, they have also been used to probe the ``missing satellite problem" \cite{Bli99,BucPet18}).
An interesting property of these objects is that they are extremely metal-poor. This is because the gravitational potential well of such galaxies is very shallow and therefore the metals produced from supernovae get efficiently ejected out of their halos \cite{McQ15}.
Their behavior of such galaxies is similar to chemically primitive galaxies in the early Universe and they can therefore be used as laboratories for studying the formation and evolution of massive stars in nearly pristine gas, and also for constraints on primordial helium abundance and effects of reionization \cite{Wei14,Gar19,Sen19,Sen19b}.
Numerous past surveys have specifically targeted such galaxies:
%relevant for missing satellite problem
THINGS \cite{Wal08_Things}, FIGGS \cite{BegChe08_Figgs}, SHIELD \cite{Can11_Sheild}, VLA-ANGST \cite{Ott12_Angst} and LITTLE THINGS \cite{Hun12_LittleThings}, among many others. Ongoing and future 21cm surveys\footnote{WALLABY \cite{Kor20_Wallaby}, MeerKAT \cite{Mad21}, Apertif \cite{van22}, FAST \cite{Zha21}, SKA \cite{SKA}}, and also optical surveys\footnote{DES \cite{Kop15_DES,Bec15_DES,Drl15_DES,Drl20}, SAGA \cite{Geh17,Mao21}, DESI \cite{DESI}, HSC \cite{Aih18}, Dragonfly \cite{Dan20}, MSE \cite{McC16}, Rubin observatory \cite{LSST,Drl19_LSST,Mut21}, Roman telescope \cite{WFIRST}} will find and characterize an even larger number of these galaxies. This motivates finding new ways of using the upcoming observations of gas-rich dwarfs to probe physics beyond the standard model.

\citet{WadFar21} (hereafter \citetalias{WadFar21}) showed that gas-rich dwarfs can be used to constrain DM models. The cooling rate of gas in such dwarfs is ultra-low, primarily because of their low metallicity (see Fig.~1 of Ref.~\cite{WadWaninprep20} for a comparison of their cooling rates with those of typical Milky Way systems). The gas in these dwarfs is therefore very sensitive to heat injection by a non-standard energy source. \WF\ used a particular well-studied gas-rich dwarf galaxy called Leo~T and required heat exchange between DM and ordinary matter to not exceed the gas cooling rate. This leads to strong limits on
DM scattering with ordinary matter and on hidden photon dark matter (HPDM). Subsequently, the gas cooling rate of Leo~T has also been used to constrain gas heating due to primordial black holes (PBHs) by Refs.~\cite{LuTak21,LahLu20,Kim20,TakLu21,Tak21b} (the gas heating from PBHs occurs via various mechanisms like radiation from the BH accretion disk, Hawking radiation, BH outflows and dynamical friction).
In this paper, we use the methodology of \WF\ to constrain the heating of gas in Leo~T caused by decay or annihilation of DM particles.

Annihilation or decay of DM can produce high-energy Standard Model (SM) particles like $\gamma$-rays and synchrotron emission.
Gas-less dwarfs (e.g., Draco, Fornax) offer relatively clean environments to probe such emissions
\cite{Col07,Bal07,Atw09,McC12,Drl15_DES2,Reg15,Reg17,Cap18,Reg20}, and their observations have set one of the strongest limits on annihilations of WIMPs \cite{Ack15}.
%Because dwarf galaxies lack nonthermal astrophysical processes, they are good targets for DM searches via $\gamma$-rays or synchrotron emission.
However, in case DM in astrophysical systems decays/annihilates to lower energy candidates like UV/soft X-ray photons, or MeV$-$GeV e$^+$/e$^-$, these are notoriously difficult to observe directly as they get scattered/absorbed by surrounding astrophysical medium \citep{Ove04,Mur20}.
Precisely in such cases, gas-rich dwarf galaxies can be a very good complement because UV photons or low energy e$^+$/e$^-$ from DM can be efficiently trapped in the gas of gas-rich dwarfs and heat it. We require such heating to be less than the ultra-low radiative cooling rate of gas in Leo~T and obtain strong constraints on DM.
It is worth mentioning that diffuse neutral clouds in the Milky Way are also astrophysical systems with low gas cooling rates. A variety of MW gas clouds have therefore been used for constraining DM-gas interactions \cite{Chi90,DubHer15,WadFar21,BhoBramante18,BhoBramante19,BhoBra20,comment}. In addition to using Leo~T, we will also use robust gas clouds from \WF\ to constrain DM.

%Ultra-faint gas-rich dwarfs are not found within the virial radius of large halos like the Milky Way because their gas gets stripped due to pressure of the surrounding ionized medium (i.e. ram pressure stripping \cite{GunGot72,Grc09, Spe14}).
%\new{Observations of such dwarfs therefore now enable a direct probe of DM-baryon interactions.}

The outline of the paper is as follows. We first discuss the properties of the Leo~T galaxy in Sec.~\ref{sec:LeoT}. We discuss heating of gas due to injection of photons and \e in Sec.~\ref{sec:injection}. Our methodology for setting bounds on DM is in Sec.~\ref{sec:DMconstraints}. We provide model independent constraints in Sec.~\ref{sec:ModelIndep}, while constraints on particular models of DM are in Sec.~\ref{sec:DMmodels}. We discuss our results and methodology in Sec~\ref{sec:Discussion} and conclude in Sec.~\ref{sec:Conclusions}.

\section{Leo~T galaxy properties}
\label{sec:LeoT}

In this section, we discuss properties of Leo~T galaxy and the reason why it is an ideal astrophysical candidate to probe heating due to decaying and annihilating DM.
Leo~T is an ultra-faint dwarf galaxy located outside the virial radius of the Milky Way (MW), nearly 420 kpc from the MW center. Leo~T is both dark-matter dominated and gas-rich, it is well-studied observationally and has garnered modelling attention \cite{faerman13,Patra18}. Ref. \cite{leoObsOld08} analyzed high-resolution Giant Meterwave Radio Telescope (GMRT) and Westerbork Synthesis Radio Telescope (WSRT) observations to determine the \HI column density and velocity profile (see also \cite{adams17}). The stellar mass and star formation history of Leo T has been studied using data from Hubble Space Telescope (HST) and Multi Unit Spectroscopic Explorer (MUSE) survey. For reference, Leo T has $M_\mathrm{HI}=4.1\times 10^5\, M_\odot$ \cite{adams17} and $M_*\sim 10^5\, M_\odot$ \cite{Zou21, weisz12}.
 
\subsection{Gas and DM density profiles}

\begin{figure}
\centering
\includegraphics[scale=0.6,keepaspectratio=true]{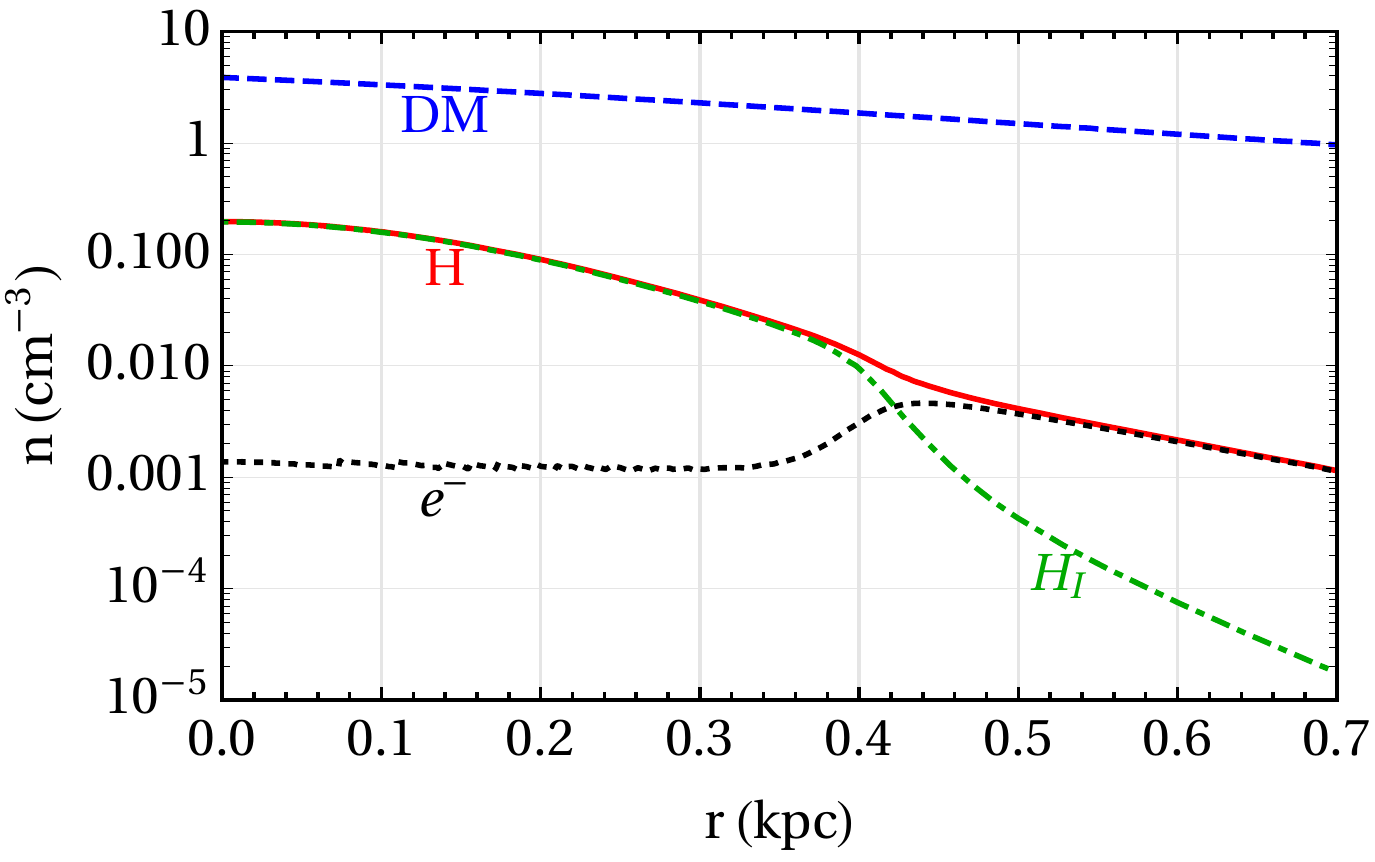}
\caption{We use a model of the gas-rich Leo~T dwarf galaxy by Ref.~\cite{faerman13}, which was fitted to 21cm measurements of the galaxy by \cite{leoObsOld08}, and is also consistent with stellar velocity dispersion estimates by \cite{Str08}.
We show the number density of DM (for $m_\mathrm{DM}=1$ GeV), atomic hydrogen (H\textsc{i}), electrons (e$^-$) and total hydrogen (H) components of the model.
This figure is taken from \citetalias{WadFar21} and is shown here for self-contained discussion.} 
\label{fig:LeoTobs}
\end{figure}

Ref. \cite{faerman13} (FSM13 hereafter) modelled the DM and gas density profiles of Leo~T. They assumed that \HI in Leo T is an isothermal gas sphere, and its DM halo has a well-motivated flat-core (Burkert) profile. They fit their model to the \HI column density and temperature measurements of Leo T by \cite{leoObsOld08}, and inferred the DM profile by assuming the gas to be in hydrostatic equilibrium. We adopt their model shown throughout this paper and show the corresponding density profiles in Fig.~\ref{fig:LeoTobs}. The DM halo mass within $r<0.35\, \textup{kpc}$ is $\sim 1.2 \times 10^7\, M_\odot$, and is consistent with dynamical mass measurements of Refs.~\cite{Bon15, Str08,leoObsOld08,SimonGeha07}.  For a more detailed discussion of the model of FSM13 and on other modeling studies of Leo~T, see Appendix A1 of \citetalias{WadFar21}. We will later discuss in section~\ref{sec:Conclusions} the sensitivity of our DM limits to uncertainities in its DM halo properties.

The electron density inside Leo~T is estimated by FSM13 with a radiative transfer code and using a model for the UV/X-ray metagalactic background. The ionization in the outermost region of Leo T is governed by the UV background flux, but that in the inner region is governed by the soft X-ray flux (as X-ray photons can penetrate the inner regions due to their have lower cross sections). While there is a large uncertainty in the strength of the
metagalactic UV flux, the X-ray flux is well constrained by current observations \cite{hm12}, and therefore the uncertainty in the electron density in the inner regions is expected to be small.

\subsection{Gas temperature}
\label{sec:GasTemp}
The temperature of gas in Leo T is directly estimated from the observed line of sight velocity dispersion of \HI ($\sigma_v$) as
\be
T_\mathrm{gas}< \left(\frac{\sigma_v}{7.1\, \mathrm{km\, s}^{-1}}\right)^2 6100\, \mathrm{K} 
\label{GasTemp}
\ee
We only have an upper limit for $T_\mathrm{gas}$ because, apart from thermal contributions, $\sigma_v$ also has contributions from turbulent or bulk motions inside the gas. Ref.~\cite{adams17} provides $\sigma_v=7.1\pm 0.4$ km s$^{-1}$ for the WNM of Leo T (see also \cite{leoObsOld08}). We conservatively use $T_\mathrm{gas}$=6100 K for the Leo T bounds in this paper, but we also report DM limits for an even more conservative estimate corresponding to the $+2\sigma$ value: $\sigma_v=7.9$ km/s, which translates to $T_\mathrm{gas}$=7552 K.

\subsection{Gas cooling rate}
\label{sec:Cooling}

%The temperature of the warm neutral medium gas in Leo~T is $\lesssim$ 6000 K \cite{adams17,leoObsOld08}.
For \HI gas with $T\lesssim 10^4$ K, collisions between hydrogen atoms are not typically energetic enough to ionize themselves. The cooling is dominated by photons corresponding to the fine-structure transition lines of metals like C, O, Si and Fe \cite{Maio07}. The gas cooling rate $\dot{C}$ depends linearly on the metal abundance as seen from the following formula \cite{grackle}:
\begin{equation}
\dot{C}=n^2_\textup{H}\, \Lambda (T) 10^\textup{[Fe/H]}\, ,
\label{eq:grackle}
\end{equation}
where $10^{[\textup{Fe}/\textup{H}]}$ is the metal fraction in the gas relative to the sun ([Fe/H] is the metallicity defined as $[\textup{Fe}/\textup{H}] \equiv \log_{10} (n_\textup{Fe}/n_\textup{H})_\textup{gas} -\log_{10} (n_\textup{Fe}/n_\textup{H})_{\textup{Sun}}$). $n_\textup{H}$ is the total hydrogen number density, and $\Lambda(T)$ is the cooling function which depends on the temperature of the gas (a rough approximation being $\Lambda(T)\simeq 10^{-27.6}\, T^{0.6}$ erg cm$^3$ s$^{-1}$ for $T\sim(300-8000)$ K, see Fig.~7 of \WF). Gas-rich low-mass dwarfs generally have ultra-low radiative cooling rates as they are metal-poor and have low $n_\textup{H}$.
In particular, Leo~T has [Fe/H] $\sim -1.74 \pm 0.04$ given by spectroscopic measurements from~\cite{Kir13} (\cite{Kim20}).

The outer part of Leo~T ($r>$ $0.35$ kpc) is ionized due to the metagalactic UV background. It is difficult to find a robust measure of the rate at which the outer ionized region cools. Therefore, we restrict our study to the warm neutral medium (WNM)  within $r<0.35$ kpc.
Compared to using Eq.~\ref{eq:grackle}, one can obtain a slightly more accurate estimate of the cooling of the gas by taking into account the ionization and density profiles from Fig.~\ref{fig:LeoTobs} and explicitly calculating the metal line transition rates (see Ref.~\cite{Kim20}, hereafter \Kim). 
We adopt the result of \Kim\ for the volume-averaged radiative cooling rate of the gas in Leo~T for $T_\mathrm{gas}\sim 6000$ K:
 $\dot{C}\simeq 7 \times 10^{-30}$ GeV cm$^{-3}$ s$^{-1}$. 
For the more conservative case of $T_\mathrm{gas}$=7552 K,
 the cooling rate becomes $\dot{C}\simeq 1.46 \times 10^{-29}$ GeV cm$^{-3}$ s$^{-1}$.
Further discussions on the gas cooling rate are in Appendix~\ref{apx:cooling}.
Note that the cooling rate can also be directly estimated using the intensity of emission or absorption of the C~\textsc{ii} line \cite{Leh04}; it would be a valuable complement to have those observational estimates for Leo~T.

Apart from Leo~T, we also use core of a diffuse neutral cloud G33.4$-$8.0 in the Milky Way \cite{PidLock15}. The parameters we use for the cloud are $n_{\textup{H}_\textup{I}}$ = $0.4\pm0.1$ cm$^{-3}$, $T$ = $400\pm90$ K and $[\textup{Fe}/\textup{H}]\sim 0$ \cite{PidLock15} (see \WF\ for further details about the cloud). We use a rough estimate for the volume-averaged cooling rate, $\dot{C} \sim 2.1 \times 10^{-27}$ erg cm$^{-3}$ s$^{-1}$, calculated using Eq.~\ref{eq:grackle}.

%---------------------------------------------------------------------
\section{DM Energy injection scenarios}
\label{sec:injection}

The energy deposition in \HI gas due to decay and annihilation of DM particles is given by 
\be\begin{split}
&\left(\frac{dE }{dt dV} \right)_\mathrm{decay} = n_{\chi} \Gamma_\chi E_\textup{heat}\\
&\left(\frac{dE }{dt dV} \right)_\mathrm{annihilation}= \frac{1}{2} n^2_\chi \langle\sigma v\rangle E_\textup{heat}
\end{split}
\label{eqn:decayInjection}
\ee
where $n_{\chi}$ is the DM number density and $\Gamma_\chi$ is the decay rate, $\langle\sigma v\rangle$ is the velocity-averaged annihilation cross section, and $E_\textup{heat}$ is the heat energy injected in the \HI gas due to interactions of the decay and annihilation products with the gas. The factor of 1/2 corrects the double counting of annihilation of DM particles.  We consider decay/annihilations of DM to either $e^\pm$ or photons in this paper. Let us now derive $E_\textup{heat}$ for the two cases separately.

\subsection{$e^\pm$ energy deposition in gas}
\label{sec:eEnergy}

We consider relativistic $e^\pm$ in the MeV$-$GeV range produced from DM in this section.
The kinetic energy of \e is lost due to collisions with free electrons/ions in the gas and in ionizing/exciting neutral atoms\footnote{A fraction of the \e energy could also be lost due to inverse Compton scattering of CMB or optical photons \cite{Val10,Gal13}. This effect is however subdominant as compared to the ionization energy loss for low energy \e (i.e. $\lesssim$ GeV) at low redshifts where the density of photons is significantly lower than the early Universe. \e can also lose their energy due to synchrotron emission. However, the synchrotron losses dominate only for very high energy \e and we have checked that this effect is negligible for the range of masses considered in our plots ($t_\mathrm{loss, sync}\sim 12$ Gyr for GeV \e).}. Only a fraction $(f_e (E))$ of the initial K.E $(E)$ is therefore converted to heat due to collisions or absorptions in the gas.
%$f_\textup{e}(E)$ is the fraction of the initial kinetic energy ($E$) of $e^+$/$e^-$ that is deposited as heat \footnote{only a fraction of the energy of decay products is converted to heat due to collisions or absorptions in the gas, and the rest is used up in ionizing/exciting neutral atoms in the gas}. 
We use the fitting function from \cite{Ric02} (which is based on an earlier work by \cite{Shu85} and is in good agreement with simulations for $E\gtrsim 11$ eV \cite{Fur10}):

\be\begin{split}
f_\textup{e}(E)\simeq\, & 1-(1-x_e^{0.27})^{1.32}\\
& +3.98 \bigg(\frac{11\mathrm{eV}}{E}\bigg)^{0.7}x_e^{0.4} (1-x_e^{0.34})^2
\label{eq:fe}\end{split}\ee
where $x_e$ is the ionization fraction of the gas. To give a rough idea, when the kinetic energy of primary electrons is beyond keV, their energy is distributed roughly equally in heating, ionization and atomic excitation of the gas \cite{Fur10}. 
%($f_e(E>1\, \textup{KeV})\sim$ 0.4).
Note that using $f_\textup{e}(E)$ is conservative since it does not include heating of gas via the dissipation of \alf waves produced by the motions of $e^\pm$ (see the discussion below for details).
The net heat deposition energy is given by 
\be
E_\textup{heat}\equiv \sum_{i\in e^{\pm}} (E_i-m_i) f_\mathrm{e}(E_i-m_i) (1-e^{-\lambda_e})
\label{eq:Eheate}
\ee
where $(E_i-m_i)$ is the kinetic energy of decay product $i$ and $\lambda_e$ is opacity of the \HI gas to \e. We use the size of Leo~T as the extent of its warm neutral medium ($r_\mathrm{WNM} \simeq 0.35$ kpc) in our opacity calculations.
Assuming the gas is comprised of H and He with primordial compositions, the opacity is
\be
\lambda_e\simeq \sum_{i=\textup{H,He}} \rho_i\, c\, t_\mathrm{confine}\, S_i(E)/E
\label{eq:eOpacity}\ee
%The rest of the energy is lost as it is carried away by secondary photons with $E<10.2$ eV as these exit the plasma without further collisions.
%There is one other effect which is important to consider: how much time do the relativistic $e^{\pm}$ spend inside the \HI medium. One therefore needs to calculate the opacity of the gas $\lambda$ given by:
where $S_i(E)$ is the electron stopping power of element $i$ taken from \cite{NIST} (for e.g., $S_\textup{H}(E= 1\, \textup{MeV})\simeq$ 3.8 MeV cm$^2$/g), $\rho_i$ is the mass density, and $t_\mathrm{confine}$ is the time spent by \e before they escape the galaxy.
A naive estimate of $t_\mathrm{confine}$ can be $r_\mathrm{WNM}/c \sim \mathcal{O}$ (kyr) assuming that the relativistic $e^\pm$ ballistically escape the galaxy at nearly the speed of light ($c$) \cite{LahLu20}.
This is however not the case because the relativistic \e stream along magnetic field lines in the galaxy at much lower speeds (see the discussion in section 5.6 of \Kim).
The situation here is similar to that of relativistic cosmic rays in the Milky Way disk; these do not escape ballistically but rather spend $\gtrsim 10^7$ yr in the disk as inferred from isotope abundance measurements \cite{Lip14,Yan01}.

One of the reasons behind \e moving slower than $c$ is that they are scattered by \alf waves excited in the plasma of Leo~T. Analogous to transverse waves propagating along a string, \alf waves are magnetohydrodynamic waves propagating along magnetic field lines, where the field strength provides tension and ions in the plasma provide inertia.
The motion of \e can excite \alf waves, which in turn react on the \e and scatter them \cite{Kul69,Ces80} (the faster the \e move, the more they get scattered due to the generated \alf waves and consequently slow down).
%due to their motion in the plasma of Leo~T. The \e scatter on the waves they themselves have generated.
%but, due to the CR streaming instability, electrons propagating in the interstellar medium scatter on self-excited \alf waves and couple to the gas (see for a diagram of the interaction). These do not move with relativistic velocities, rather with the much smaller 
Their streaming velocity along field lines is much smaller than $c$ and is close to the ion
\alf velocity (see Appendix~\ref{StreamingVel} for further details):
\be
v_A\equiv \frac{B}{\sqrt{4 \pi \rho_\mathrm{ion}}} \sim 63\, \mathrm{km/s}\, \frac{B}{1\, \mu \mathrm{G}} \left(\frac{\rho_\mathrm{ion}}{ 10^{-3}\, \mathrm{GeV/cm}^3}\right)^{-1/2}
\label{eq:vA}\ee

If the field lines were rigid and oriented in the radially outward direction, the $e^\pm$ would stream straight along the field lines at $v_A$ and the time to escape would be $r_\mathrm{WNM}/v_A \sim 5.4$ Myr. However the field lines are likely tangled rather than straight, and are also constantly being rearranged by turbulence. Thus even though \e stream along field lines, the motion of \e resembles diffusion as the field lines are themselves in a random walk configuration (see e.g.~\cite{Kru20}). We discuss these effects further in Appendix~\ref{FieldDiffusion} and use the conservative estimate $t_\mathrm{confine}\simeq 19$ Myr from Eq.~\ref{tconfine}. We show the opacity of Leo~T as a function of \e energy in the bottom panel of Fig.~\ref{fig:opacity}. The opacity for the MW co-rotating clouds is discussed in Appendix~\ref{apx:opacity}.

The calculations performed till now assume that e$^\pm$ are closely tied to magnetic field lines. To verify this, we calculate their relativistic gyration radius as
\be
r_g\equiv \frac{\gamma m_e c}{q B}=5.4\, \gamma\, \frac{1\, \mu\mathrm{G}}{B} \times 10^{-10}\, \mathrm{pc} 
\ee
which is indeed much smaller than the expected coherence length of the magnetic field ($\gtrsim$1 pc) in galaxies.
Let us now discuss the magnetic field strength in Leo~T ($B$). 
While there is no direct measurement of $B$, one can make reasonable estimates for its the largest and smallest probable values.
$B$ should be much larger than the value of the extragalactic field strength (which is expected to be $\sim 1$ nG \cite{Bla99}). This is because the magnetic field in Leo~T is enhanced by turbulent dynamo processes involved in star formation inside the galaxy.
%and via gas collapse in its gravitational potential well.
%(SFR $\sim 10^{-5} M_\odot$/yr \cite{weisz12})
The upper limit of $B$ can be inferred by comparing the star formation rate (SFR) in Leo~T with other dwarfs.
Strong magnetic fields ($>6 \mu $G) are observed only in dwarfs with extreme characteristics such as starburst dwarfs with much higher metallicity and global SFR than typical local group dwarfs \cite{Chy11}. Leo~T on the other hand has very low SFR $\sim 10^{-5} M_\odot/$yr \cite{weisz12} and therefore $B$ should be lower. One can also make a rough indirect estimate of $B$ based on extrapolation of the calibrated relation between $\Sigma$SFR and the strength of magnetic field in dwarf irregular galaxies from Ref.~\cite{Chy11}, which gives $\lesssim 1 \mu$G for Leo~T.
%The magnetic field estimate from the equipartition theorem is $\sim 1\, \mu$G \citepalias{Kim20}.
We adopt the conservative value of $B=1\mu$G in our calculations for the gas opacity and note that lower values of $B$ will lead to longer confinement timescales (because of smaller $v_A$) and therefore stronger limits on DM.

%In this paper, we only consider heating of gas due to relativistic \e produced by DM within the WNM region. However, the extent of the DM halo is much larger the WNM region. Therefore, a significant number of relativistic \e can be produced outside of the WNM and diffuse into the WNM and then lose their energy due to collisions with H, He atoms. We leave a study of this additional heating contribution to a future work.

%(see also Sec.~5.6 of Ref.~\cite{Kim20} for an alternate derivation but with comparable results).

\subsection{Photon energy deposition in gas}

Photons can transfer energy to gas via different mechanisms like the photoelectric effect, compton scattering, pair production, Rayleigh scattering, and photonuclear absorption (see Fig~34.15 of \cite{PDG}).
For low energy photons (i.e. $E_\gamma\lesssim$100 keV, which is the range we are most interested in for this paper), photons primarily heat the gas via the photoelectric effect where they ionize H or He atoms in the \HI gas.
A fraction of the photon energy in that case is spent in the ionization of the atoms and the rest is converted into the K.E of the electron. The energy finally deposited as heat in the gas is therefore
 \be
E_\textup{heat}\simeq \sum_i (E_i-I_H) f_\mathrm{e}(E_i-I_H) (1-e^{-\lambda_\gamma})
\label{eq:Eheatgamma}
\ee

\begin{figure}
\includegraphics[scale=0.55,keepaspectratio=true]{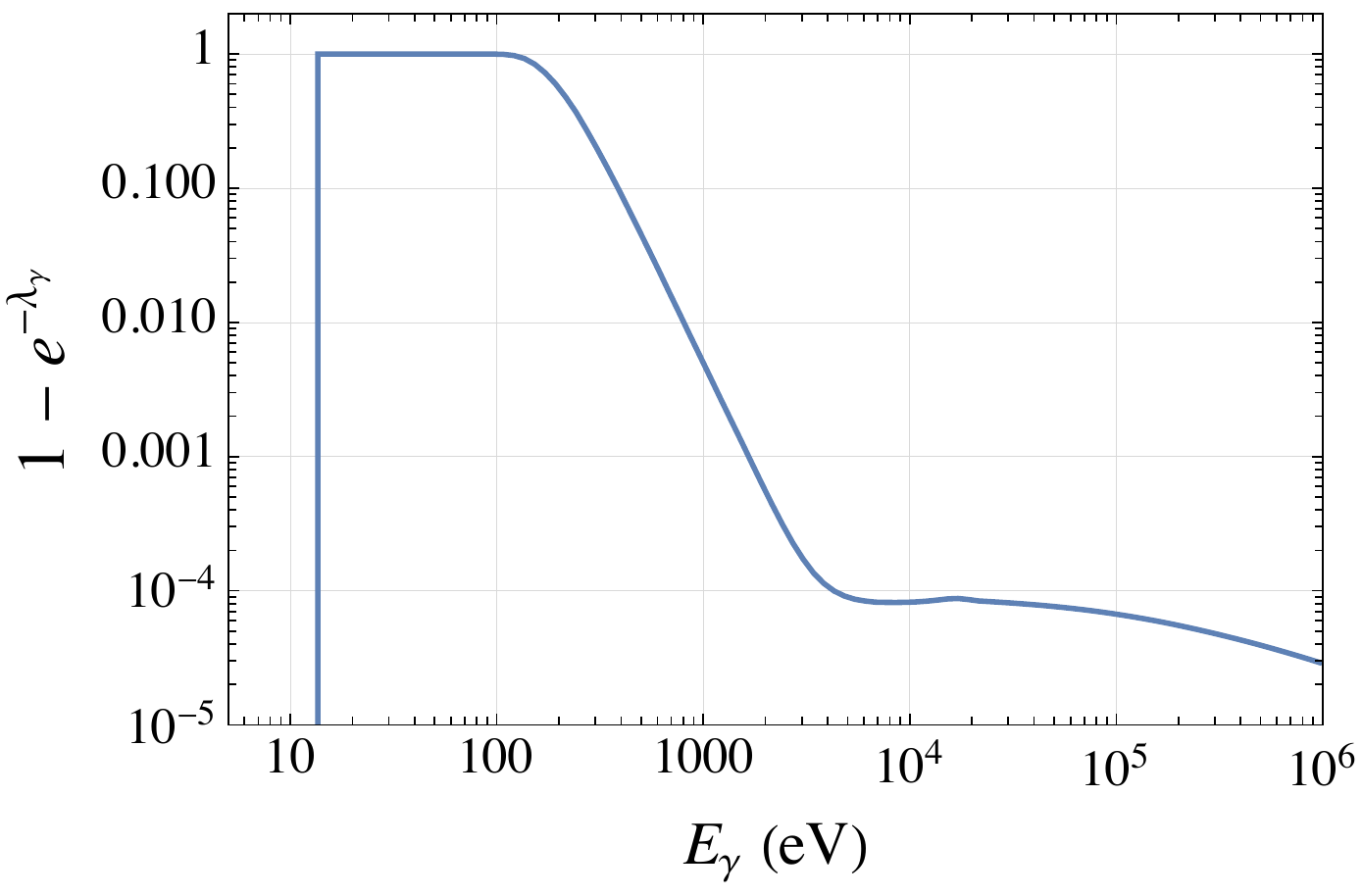}
\includegraphics[scale=0.53,keepaspectratio=true]{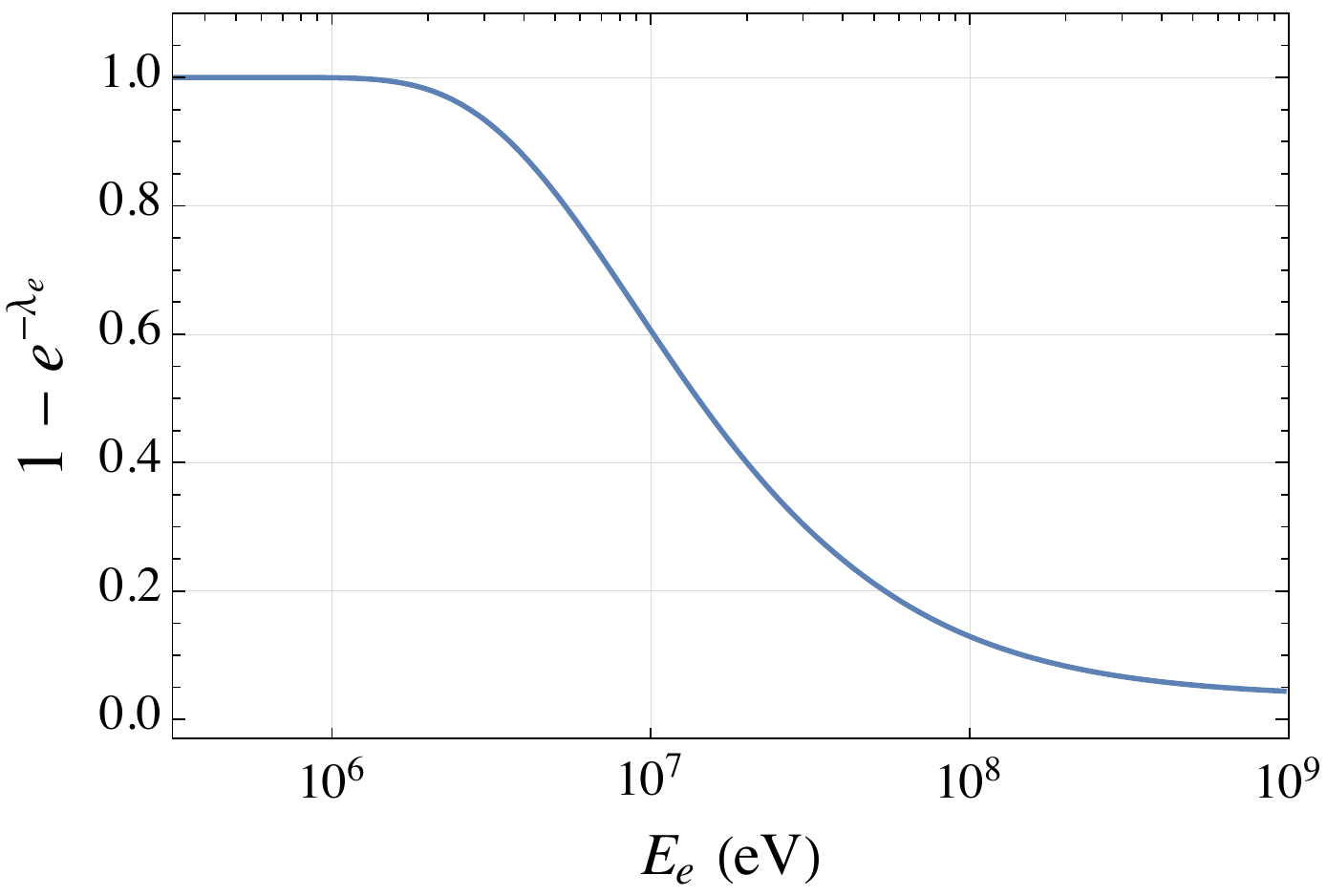}
\caption{Opacity factor of the \HI gas in Leo~T as a function of energy of injected photons (top) and kinetic energy of injected $e^+/e^-$ (bottom). We use Eq.~\ref{eq:photoOpacity} for $\lambda_\gamma$ and Eq.~\ref{eq:eOpacity} for $\lambda_e$ (we use $B$ 1$\mu$G and $t_\mathrm{confine}\simeq$ 19 Myr for this plot). Leo~T is a strong probe of photons with $E_\gamma\lesssim200$ eV and electrons with $E_e\lesssim 10$ MeV.} 
\label{fig:opacity}
%\vspace{-0.1in}
\end{figure}
%We consider photons heating the gas due to photoelectric effect.
%Note that for energies $\gtrsim$100 keV, photons can also lose significant amounts of energy due to compton scattering, Rayleigh scattering, and photonuclear absorption . 
%In this paper, we conservatively only consider the photoelectric heating. For photons interacting with the \HI gas, a fraction of their energy is spent in the ionization of the atoms and the rest is converted into the K.E of the electron. 
where $I_H$ is the ionization energy of Hydrogen and $f_e(E_i-I_H)$ is again the heat deposition fraction of the K.E of the electron from Eq.~\ref{eq:fe}. The optical depth is given by
\be
\lambda_\gamma \equiv \sum_{i=\mathrm{H,He}} n_i \sigma_i(E) r_\textup{WNM}\, ,
\label{eq:photoOpacity}\ee
which takes into account the photon fraction which escapes the galaxy without collisions with the gas and is shown in Fig.~\ref{fig:opacity}. $\sigma(E)$ corresponds to effective photon cross section\footnote{For $E_\gamma<30$ eV, we use the following fitting function from Ref.~\cite{Ban90}: $\sigma(E) = \sigma_0 y^{-3/2} (1+y^{1/2})^{-4}$, where $y\equiv E/E_0$ and $\{E_\mathrm{ion}/\mathrm{eV},\, \sigma_0/(10^{-18}\mathrm{cm}^2),\, E_0/\mathrm{eV}\}$ is \{13.6,606,6.8\} for H and \{24.8,3.94,112.88\} for He.\\
For $E_\gamma>30$ eV, we use the attenuation length data from \cite{Hen93,PDG}:\\
\href{https://henke.lbl.gov/optical_constants/atten2.html}{\textcolor{blue}{https://henke.lbl.gov/optical\_constants/atten2.html}}\\
\href{https://physics.nist.gov/PhysRefData/XrayMassCoef/tab3.html}{\textcolor{blue}{https://physics.nist.gov/PhysRefData/XrayMassCoef/tab3.html}}} including all the effects mentioned at the beginning of this sub-section, and $r_\mathrm{WNM} = 0.35$ kpc. The opacity of \HI in Leo~T is shown in Fig.~\ref{fig:opacity}. The decline seen below $ E_\gamma \lesssim$ 5 keV is because the photo-electric cross section gets significantly damped: $\sigma_\mathrm{PE}\propto E_\gamma^{-3.5}$, while it levels out for $E_\gamma \gtrsim$ 5 keV due to Compton scattering and pair production contributions. We also assume \HI gas is optically thin to radiation below the ionization threshold of 13.6 eV\footnote{Note that one exception for this case is for a very small region of photon energies close to 10.2 eV (which corresponds to the Ly$\alpha$ line), where photons are resonantly absorbed by \HII. Furthermore, dust grains in the gas can also absorb photons lower than 13.6 eV \cite{WolMckee03}.
We do not consider this additional photoionization heating effect so our limits are conservative.}.

It is important to emphasize that the gas in Leo~T is optically thick for 13.6$<E_\gamma\lesssim200$ eV (i.e. the extreme UV [EUV] range), which is precisely the range where constraints from direct measurement of photon backgrounds are the weakest (see Fig.~1 of \cite{Ove04}).
EUV background measurements are notoriously challenging because EUV photons are strongly scattered/absorbed in the ISM (we currently only have upper bound estimates for the EUV spectrum). As we will see later in Figs.~\ref{fig:ModelIndep} and~\ref{fig:Axion}, Leo~T is highly complementary to other astrophysical measurements of photons in the EUV range.
% Photons in the UV range are efficiently absorbed by \HI gas and therefore DM decaying or annihilating to UV photons can be constrained by Leo~T. 

%------------------------------------------------------------------------------------------
\section{Constraints on DM}
\label{sec:DMconstraints}

% The heating efficiency of electrons and photons on the gas cloud in Leo~T is discussed in detail in the previous section. With this knowledge, we can directly compute the heating rate due to these decay and annihilation channels and set constraints on the decay lifetime and annihilation cross section.
In this section, we derive our constraints on annihilating/decaying DM, both for model-independent scenarios and for particular models of DM.
Let the volume-averaged rates: \{$\dot{C},\dot{H}$, $\dot{Q}$\} correspond to astrophysical cooling, astrophysical heating and DM heating respectively. For a system to be in a steady state, we need $|\dot{Q}| = |\dot{C}-\dot{H}|$. In this paper, we set conservative bounds on the DM interaction cross section by requiring $\dot{Q}\leq \dot{C}$. Note that more stringent bounds could in principle be placed by including the astrophysical heating rate $\dot{Q}\leq |\dot{C}-\dot{H}|$.

We use
$\dot{Q} \equiv \frac{1}{V} \int dV\, \frac{dE }{dt dV} = \frac{1}{r^3_\mathrm{WNM}/3} \int_0^{r_\mathrm{WNM}} dr\, r^2 \frac{dE }{dt dV}$
where we substitute $dE/dtdV$ from Eq.~\ref{eqn:decayInjection}.
% the only radial dependent piece in $dE/dtdV$ is the DM number density $n_\chi$ (see Fig.~\ref{fig:LeoTobs})
As discussed in Sec.~\ref{sec:Cooling}, we restrict the integration range to be within the warm neutral medium ($r_\mathrm{WNM} = 0.35$ kpc) for all our volume averaged integrals.
% because outside this region the gas is highly ionized

\subsection{Model independent constraints}
\label{sec:ModelIndep}
In this sub-section, we set limits on the decay lifetime of DM ($\tau^\mathrm{min}$) and on the velocity averaged DM annihilation cross section $\langle \sigma v \rangle$. We consider both $e^+ e^-$ and $\gamma \gamma$ channels.
For $\chi \to$ \e, we use Eq.~\ref{eq:Eheate} with $E_e = m_\chi/2$ (for $\chi\chi \to$ \e, we instead use $E_e = m_\chi$), similarly also for photon energies $E_\gamma$. We show the results for all the cases in Fig.~\ref{fig:ModelIndep}. Bounds from Leo~T are shown in solid black whereas the bounds from the MW gas clouds are shown in blue. For the DM decay cases, the sketched/filled regions show the ruled-out parameter space (we do not however sketch/fill regions for the annihilation case in order to avoid confusion because two different scenarios are compared in the same plot: s-wave and p-wave). Let us discuss each of the sub-panels of Fig.~\ref{fig:ModelIndep} separately:

\begin{figure*}[t]
\begin{tabular}{ll}
\includegraphics[width=0.47\textwidth,keepaspectratio=true]{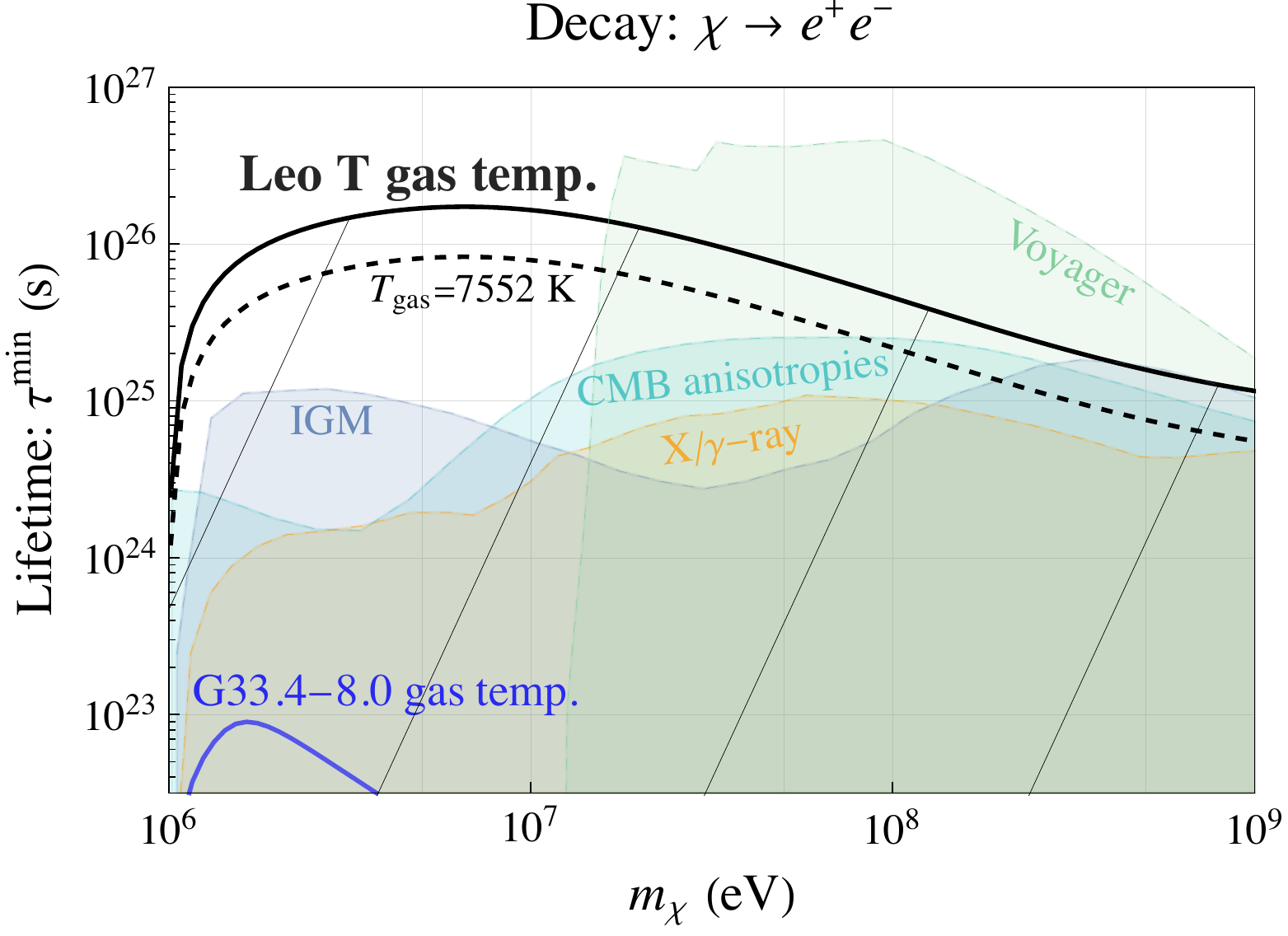}
&
\includegraphics[width=0.47\textwidth,keepaspectratio=true]{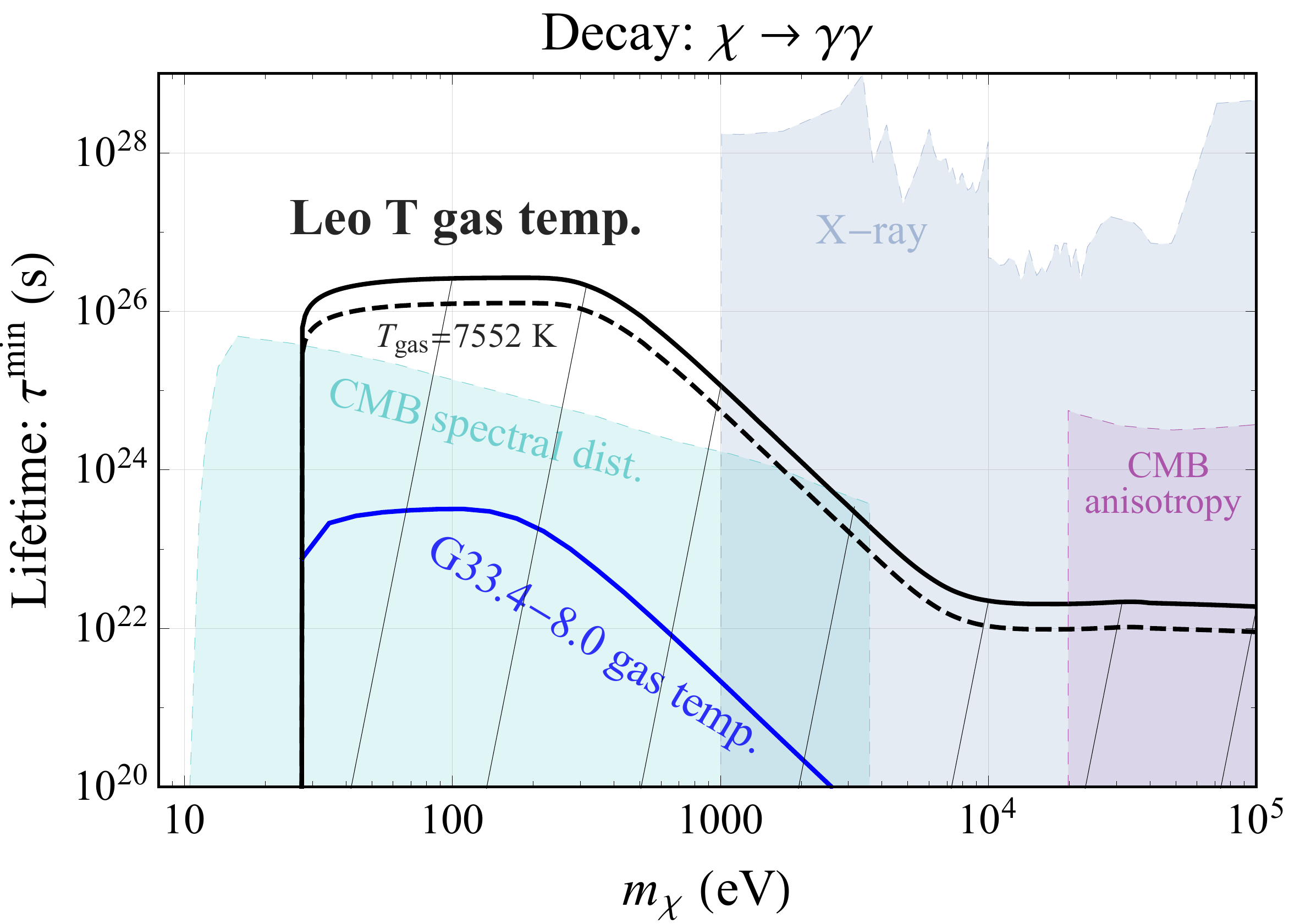} \\
\includegraphics[width=0.47\textwidth,keepaspectratio=true]{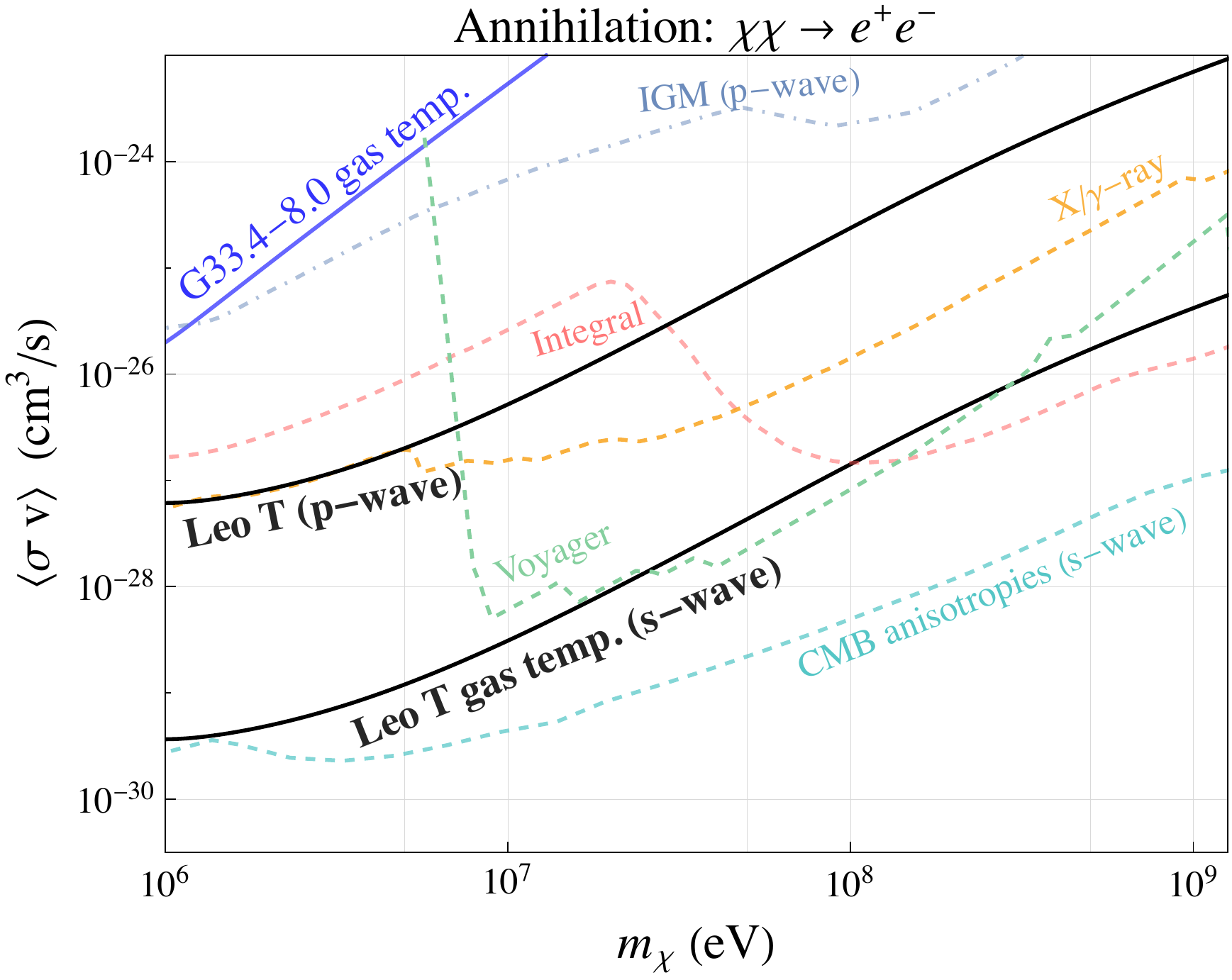} & 
\includegraphics[width=0.47\textwidth,keepaspectratio=true]{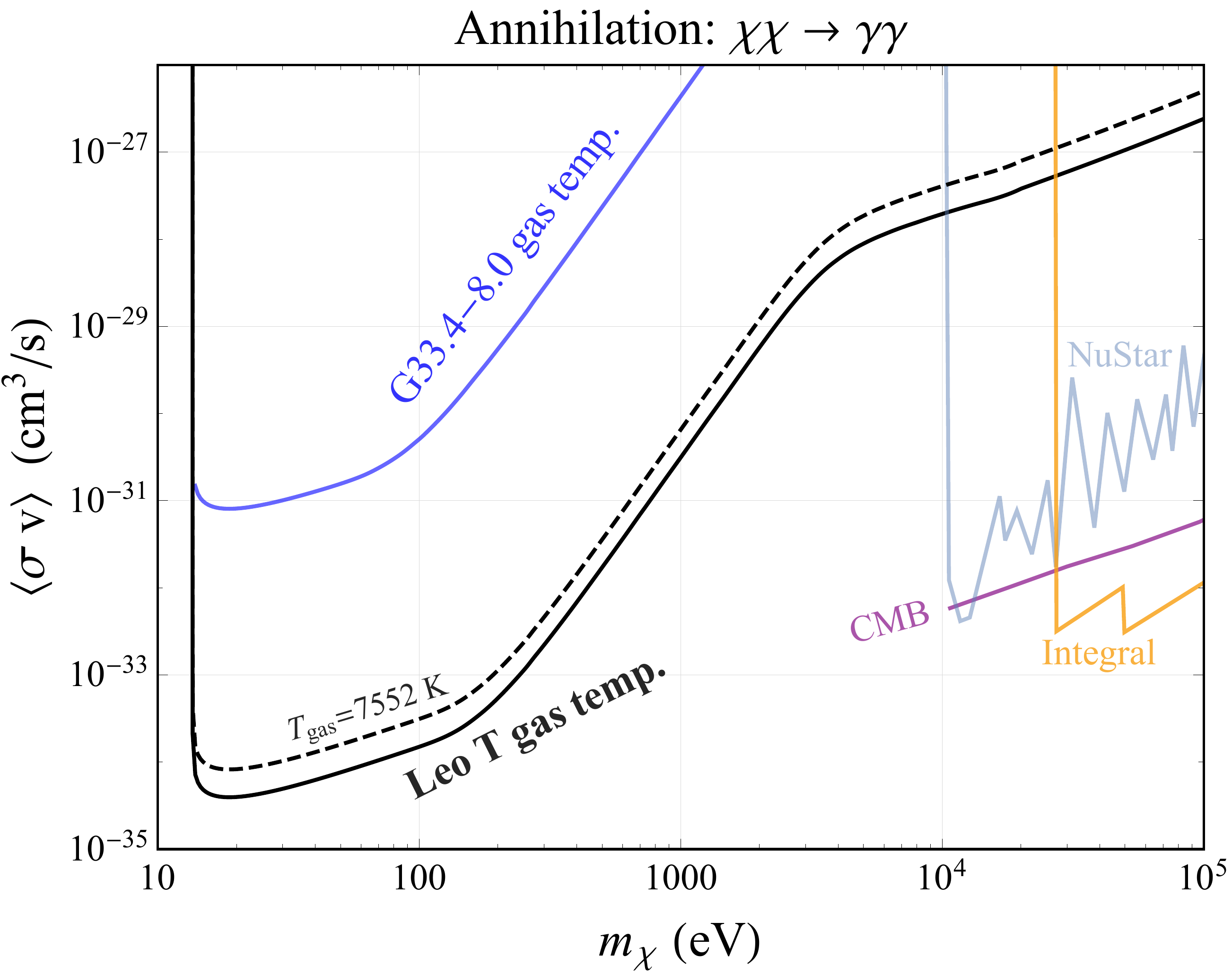}
\end{tabular}
\caption{\textit{Top left (right)}: Lower-limits on lifetime of DM decaying to $e^{\pm}$ ($\gamma\gamma$). \textit{Bottom left}: Upper-limits on s-wave and p-wave velocity-averaged annihilation cross section of $\chi\chi\to e^+ e^-$ (\textit{Bottom right}: only s-wave limits on $\chi\chi\to \gamma\gamma$). Limits from the gas-rich Leo~T dwarf galaxy are in black, and from the Milky Way co-rotating gas cloud G33.4$-$8.0 are in blue. We adopt $v_\mathrm{ref}=220$ km/s for showing the p-wave constraints. See the main text for other bounds in the figures. We also show the effect of changing the observed temperature estimate from $T_\mathrm{gas}$=6100 K to the $2\sigma$ conservative value: $T_\mathrm{gas}$=7552 K (see section~\ref{sec:LeoT}); this leads to shifting all the Leo T bounds in this figure by a factor of $\sim$2, as shown by the dashed black lines. Overall, Leo~T updates existing limits on $\chi \to$ \e lifetime by more than an order of magnitude for $m_\chi \in$ [1$-$10 MeV]. Leo~T also gives stronger constraints than all the previous literature on $\chi \to\gamma\gamma$ for $m_\chi \in$ [26.2 eV$-$keV].}
\label{fig:ModelIndep}
\end{figure*}

\begin{itemize}

\item $\chi\to$ \e: For $m_\chi\lesssim 10$ MeV, the temperature of the gas in Leo~T updates existing limits by more than an order of magnitude and sets the most stringent lower bound on $\taumin$. This has implications for the model building of MeV-scale DM~\cite{Green:2017ybv}, together with complementary probes from BBN and CMB, among others~\cite{Sabti:2019mhn,Sabti:2021reh,Giovanetti:2021izc}.
Additional limits are from the temperature of intergalactic medium (IGM)~\cite{Liu21} in gray, CMB anisotropy~\cite{Sla16} in cyan, the Voyager satellite~\cite{Boudaud:2016mos} in green, and X/$\gamma$-ray searches~\cite{Ess13} in orange. 

\item $\chi\to\gamma\gamma$: 
The Leo~T limit is seen to provide refined constraints for 30 eV$<m_\chi<$ 1 keV, and we will discuss the implication on specific DM models in Sec.~\ref{sec:DMmodels}. Additional limits are from CMB spectral distortions~\cite{BolChl20} in cyan, CMB anisotropy~\cite{Sla16} in violet, and X/$\gamma$-ray searches~\cite{Ess13,Gew17,Boyarsky:2018tvu,Foster:2021ngm} in gray.\footnote{Recent and future X/$\gamma$-ray searches can improve the previous X/$\gamma$-ray limits by about one order of magnitude for $m_\chi > 10$ keV~\cite{Ng:2019gch,Laha:2020ivk,Ando:2021fhj}, which are not included in our plot.} Note that the robust general constraint on decaying DM (dDM)~\cite{Audren:2014bca,Poulin:2016nat} independent of final state gives $\taumin \geq 4.5\times 10^{18}$ s, about 10 times the age of the Universe. As this limit is a few orders of magnitude weaker, we have not shown it explicitly in Fig.~\ref{fig:ModelIndep}.

\item $\chi\chi \to e^\pm$: Additional limits are from the Planck CMB survey \cite{Sla16} (see also \cite{Kaw21}), compilation of X-ray and $\gamma$-ray data \cite{Ess13} surveys like Integral \cite{Cir21} and Voyager~\cite{Boudaud:2016mos}, and the temperature of IGM \cite{Liu21}. Note that Leo~T gives stronger constraints for the s-wave annihilation case as compared to all the Milky Way probes. For calculating the p-wave constraints, we adopt $v_\mathrm{ref} = 220$ km/s corresponding to the typical velocity distribution observed in the MW (this involves rescaling the Leo~T s-wave limits by a factor (7$\sqrt{6}$/220)$^2$, given the 1D velocity dispersion of DM is $\sim$7 km/s \cite{leoObsOld08,SimonGeha07,adams17}). It is worth noting at beyond $m_\chi>$ GeV, there are strong limits from $\gamma$-ray observations of gas-less dwarf spheroidals \cite{Ack15, Alb17} and of extragalactic halos \cite{Lis18}.

\item $\chi\chi \to \gamma\gamma$: Additional limits are from CMB~\cite{Liu:2016cnk}, NuStar~\cite{Ng:2019gch} and Integral~\cite{Laha:2020ivk}. Note that, to the best of our knowledge, the only probes in the sub-keV regime are provided by Leo~T and MW gas clouds. The spectral distortions of the CMB can in principle constrain the sub-keV regime, but the existing analyses have focused on the limits for DM decay rather than annihilation. The calculation for recasting the decay limits from CMB to the annihilation case is highly non-trivial hence we do not attempt it in this paper.
%essentially we obtain the first constraints on sub-keV DM annihilation cross sections.

\end{itemize}

% For the $e^+e^-$ decay final state, we combine Eq.~\ref{eqn:decayInjection} and Eq.~\ref{eq:Eheate} with $E_e = m_\chi/2$ to obtain the heating rate. Similarly, the heating rate due to $\chi\to\gamma\gamma$ is given by Eq.~\ref{eqn:decayInjection} and Eq.~\ref{eq:Eheatgamma} with $E_\gamma = m_\chi/2$.

%The Leo~T lower bound on the lifetime of DM decay to $e^{\pm}$ ($\gamma\gamma$) shown as the solid black curve in the top left (right) panel of Fig.~\ref{fig:ModelIndep}. The analogous limits from gas clouds in MW are depicted by the blue curves.
% \begin{figure}
% \centering
% \includegraphics[width=0.47\textwidth,keepaspectratio=true]{figures/eDecay.pdf}
% \includegraphics[width=0.47\textwidth,keepaspectratio=true]{figures/photonDecay.pdf}
% \caption{\textbf{Top}: Bounds on lifetime of DM decaying to $e^{\pm}$. \textbf{Bottom}: Bounds on lifetime of DM decaying to $\gamma\gamma$.}
% \label{fig:ModelIndep}
% \end{figure}

%\subsection{s-wave and p-wave DM annihilation}
% Apart from decays, we consider the case of annihilation of DM to $e^\pm$ or two photons. The corresponding limits are shown in the bottom left and right panel of Fig.~\ref{fig:ModelIndep} for $e^\pm$ and two photons respectively.

Throughout this work, we have assumed that all of DM is either decaying or annihilating, but one can also consider scenarios where only a fraction of DM ($f_\textup{DM}$) is undergoing these effects. Our bounds on the lifetime in Fig.~\ref{fig:ModelIndep} will simply be weakened by a factor of $f_\textup{DM}$ ( $f^2_\textup{DM}$ for the annihilation cross section). Note that this is different from the case of \WF, where the scaling with $f_\textup{DM}$ is non-trivial because the distribution of DM itself changes as a result of interactions with ordinary matter.
% DM comprises $\chi$ particles alone. In general, if the fraction of $\chi$ to the total energy density of DM is $f$, the number density of $\chi$ is scaled by $f$ and the heating rate due to $\chi$ decay is simply scaled accordingly. Therefore, the lower limit on the lifetime of $\chi$ will be weakened by a factor $f$. That said, the temperature of Leo~T gas clouds remains a robust and valid probe of decaying $\chi$ with arbitrary energy density fraction $f$. This is different from the scenario discussed in~\cite{WadFar21}, where the heat transfer relies on $f \sim \order{1}$.

%------------------------------------------------------------------------------------------
\subsection{Constraints on particular DM models}
\label{sec:DMmodels}

\subsubsection{Axions/Axion like particles(ALPs)}

\begin{figure}[tb]
\centering
\includegraphics[scale=0.55,keepaspectratio=true]{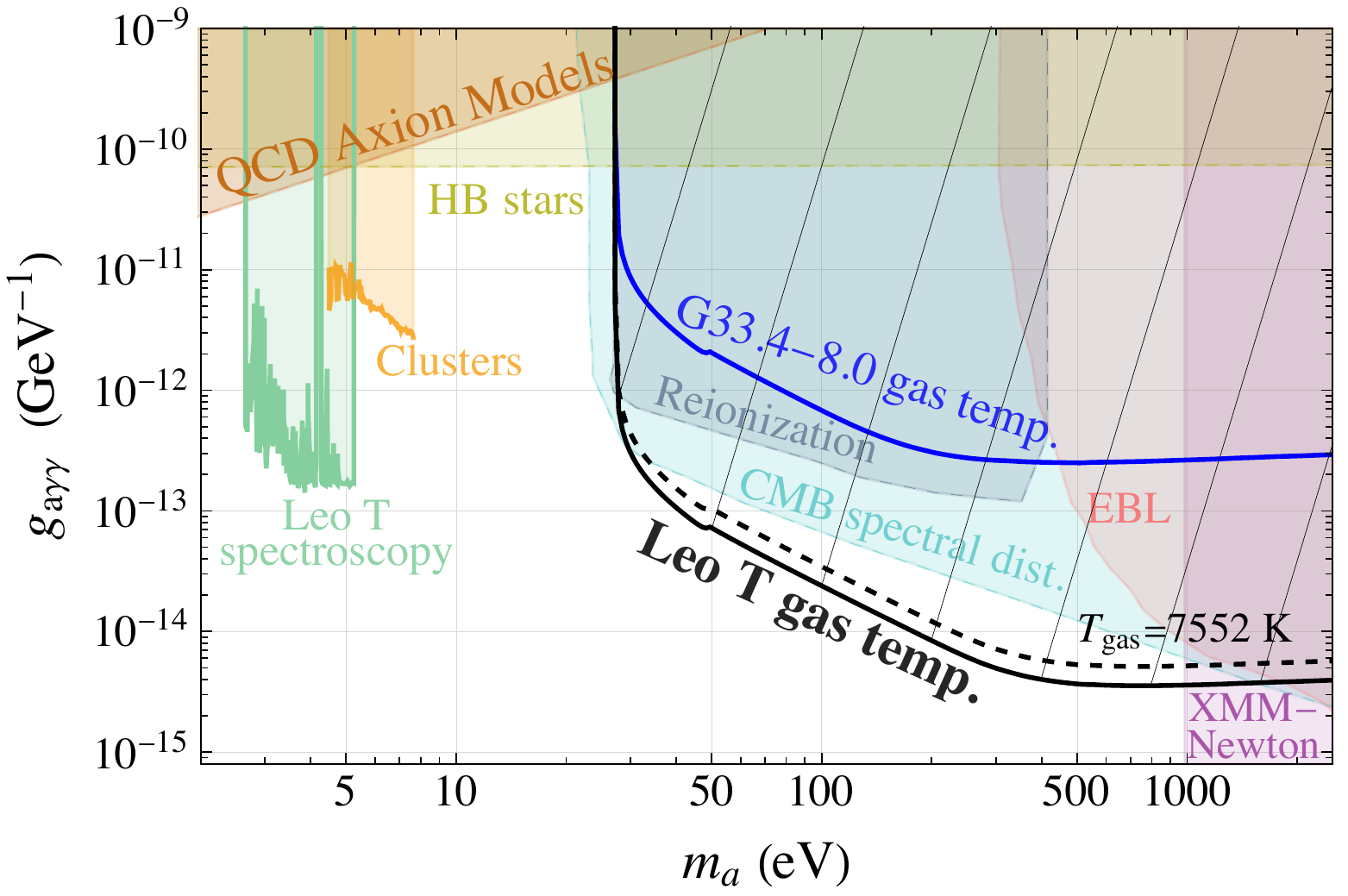}
\caption{Bounds on ALP DM with mass $m_a$ and ALP-photon coupling $g_{a\gamma\gamma}$ from the Leo~T dwarf galaxy (black) and the MW gas cloud, G33.4$-$8.0 (blue); see the text for other bounds. The dashed line shows the Leo T limit using the $2\sigma$ conservative gas temperature estimate. We also show couplings preferred by the QCD axion \cite{DiL17}. Note that the flux of photons in the extreme UV range and soft X-ray range ($E_\mathrm{photon}\sim$ 13.6$-$500 eV) is notoriously difficult to probe in astrophysical searches as these are readily scattered/absorbed in the ISM. This is precisely the range where Leo~T is the most sensitive, which makes Leo~T highly complementary to direct astrophysical searches.}
\label{fig:Axion}
\end{figure}

A widely studied example of DM that can decay to two photons is the axion or more generally, axion-like particle (ALP); see, e.g.~\cite{PQ,Cad12,Marsh:2015xka}. We consider the following generic Lagrangian of the ALP
\begin{equation}
    \lagr_a \supset -\frac{1}{2}m_a a^2 - \frac{g_{a\gamma\gamma}}{4} a F^{\mu\nu}\tilde{F}_{\mu\nu},
\end{equation}
and remain impartial to the detail of how the ALP-photon coupling is generated. The decay rate of $a\to\gamma\gamma$ can be directly computed to be
\be
\Gamma_a = \frac{m_a^3 g^2_{a\gamma\gamma}}{64 \pi}.
\ee
We recast the limits from DM decay lifetime in Fig.~\ref{fig:ModelIndep} to $g_{a\gamma\gamma}$, and show the results in Fig.~\ref{fig:Axion}. Other limits displayed in the figure include MW gas clouds in blue, CMB spectral distortions~\cite{BolChl20} in cyan, reionization~\cite{Cad12} in gray, XMM-Newton \cite{Gew17} in violet, horizontal branch stars~\cite{Raffelt:1990yz} in tan, extragalactic background light (EBL)~\cite{Cad12} in red, optical searches in galaxy clusters Abell 2667 and 2390~\cite{Grin:2006aw} in light orange, and finally spectroscopic measurements of Leo~T from the MUSE spectrograph~\cite{Reg20} in green. The range of $g_{a\gamma \gamma}$ for QCD axion models~\cite{DiL17} is also shown as the orange band in the figure. We have not shown forecasts from future line intensity mapping surveys~\cite{Ber21} for $m_a \lesssim 1$ eV and X-ray surveys (e.g., eROSITA, ATHENA \cite{Ner16}) for $m_a \gtrsim 1$ keV. Currently, Leo~T provides the best limit on ALP-photon coupling for $m_a\sim$ 30 eV$-$keV. 
It is important to mention that our bounds put significant pressure on keV-mass axion explanations of the XENON1T anomaly \cite{Apr20}, even if these axions constitute a very small fraction of the DM.

Before proceeding, let us briefly comment on another heating mechanism due to ALPs. So far we have assumed all the ALPs are from the DM halo. A DM independent source of ALPs could be stars and supernovae~\cite{Raffelt:1990yz}. This is because the coupling of ALPs to photons and/or charged fermions allows thermal production of ALPs in stellar environments. While these ALPs propagate in the Universe, they can be converted to photons near magnetic fields, or just spontaneously decay to photons and/or charged fermions~\cite{Calore:2020tjw}, and thus inject energy to astrophysical systems. Setting limits on this scenario by Leo~T requires a detailed model of the flux of ALPs and the propagation effects, which is beyond the scope of the present paper.

%------------------------------------------------------------------------
\subsubsection{Higgs portal scalars}

New CP-even scalars have extensively been studied recently in many different contexts of DM models, such as Higgs portals~\cite{McDonald:1993ex}, dark Higgs~\cite{Mondino:2020lsc}, dark glueballs~\cite{Soni:2017nlm}, and many others~\cite{Croon:2019ugf,Cheng:2021kjg}. In this section, we follow Ref.~\cite{Dev:2020jkh} and consider a generic scenario of a keV-scale scalar $S$ that mixes with the SM Higgs by an angle $\theta$. A simple realization of the scenario may be achieved by a Higgs portal operator $\lambda |H|^2 S^2$ with both $H$ and $S$ acquiring a vacuum expectation value (VEV), and then the mixing angle $\theta$ would be determined by the coupling constant $\lambda$ and the two VEVs. More concrete examples of the realization are given in~\cite{Dev:2020jkh}, where, for particular models, it is shown that the desired DM abundance can be established by the freeze-in mechanism if $\theta \sim 10^{-16}$. 

The $S$-Higgs mixing leads to interactions between $S$ and SM particles. For $m_S <2m_e$, the dominant decay channel of the scalar is $S\to \gamma\gamma$ through a loop of charged fermions similar to the Higgs. The decay rate is then given by~\cite{Dev:2020jkh}
\begin{equation}
    \Gamma_{S\to\gamma\gamma} = \frac{121}{9} \frac{\alpha^2 m_S^3 \theta^2}{512\pi^3 v_{\mathrm{EW}}^2} = 1.13\times10^{-6} \qty(\frac{m_S}{1\,\mathrm{keV}})^3 \theta^2\, \mathrm{s}^{-1},
    \label{eqn:Sdecay}
\end{equation}
where $\alpha=1/137$ is the electromagnetic fine structure constant, and $v_{\mathrm{EW}}=246$ GeV is the electroweak VEV. 

\begin{figure}[tb]
\centering
\includegraphics[width=0.45\textwidth,keepaspectratio=true]{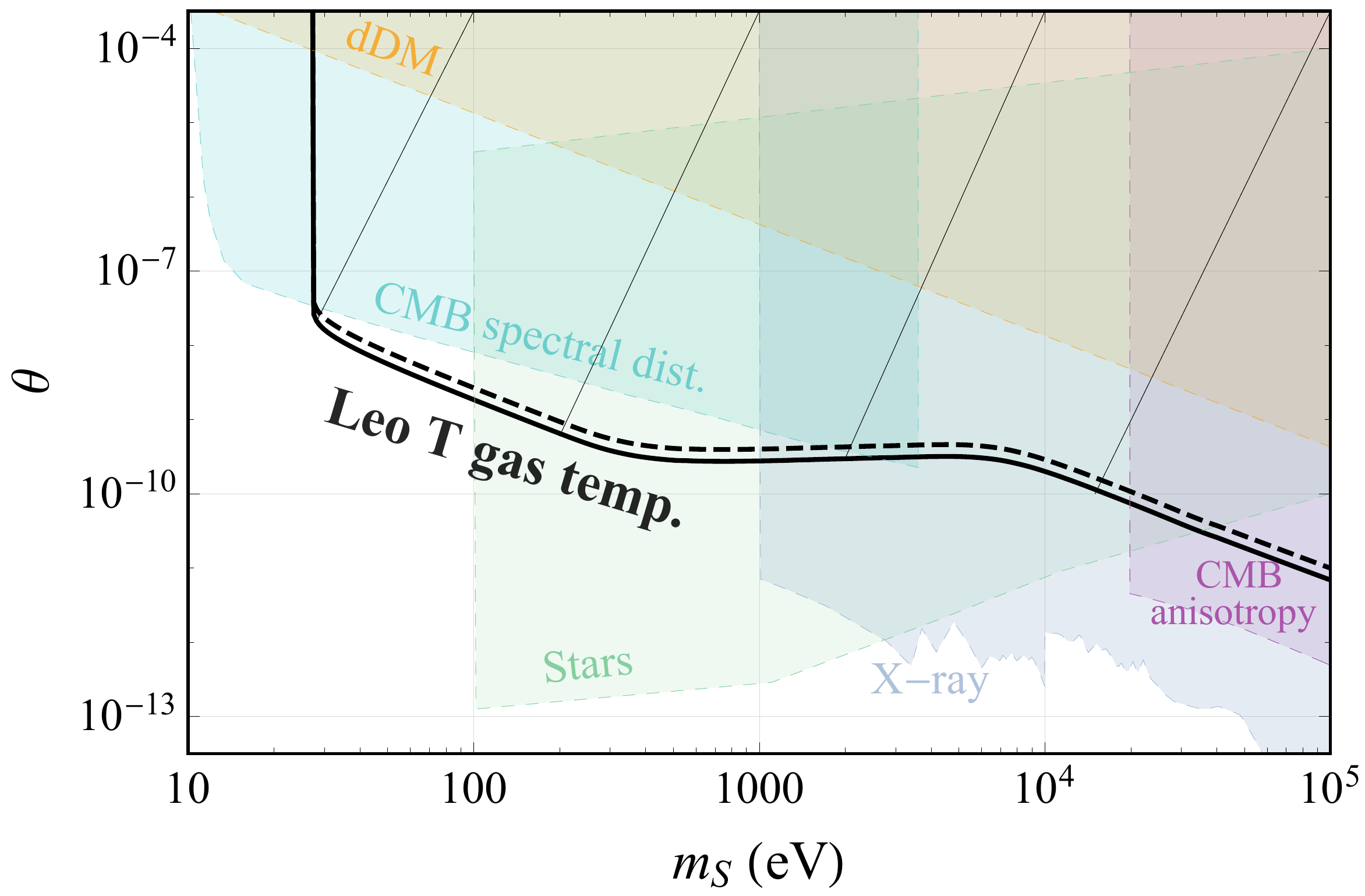}
\caption{Upper limits on the mixing parameter $\theta$ between the scalar $S$ and Higgs as a function of $m_S$. Bounds are from Leo~T (black), general dDM constraints~\cite{Audren:2014bca,Poulin:2016nat} in orange, stellar luminosity limits~\cite{Dev:2020jkh} in green, X-ray searches~\cite{Ess13,Gew17,Boyarsky:2018tvu,Foster:2021ngm} in gray, CMB spectral distortions~\cite{BolChl20} in cyan, and CMB anisotropy~\cite{Sla16} in violet. The dashed line shows the Leo T limit using the more conservative gas temperature estimate (7552~K instead of 6100~K).}
\label{fig:scalarLimits}
\end{figure}

All constraints on the decay lifetime of DM to two photons from Fig.~\ref{fig:ModelIndep} can be used to constrain $S\to \gamma\gamma$. The resulting limits on $\theta$ are shown in Fig.~\ref{fig:scalarLimits}. An additional limit is the stellar limit on $S$ shown by the light green region, which is independent of requiring $S$ to be DM~\cite{Dev:2020jkh}.\footnote{We have combined solar, red giant, and horizontal branch limits from Ref.~\cite{Dev:2020jkh}, where limits from white dwarfs are also derived and are shown to be stronger. However, the white dwarf limit is sensitive to the luminosity of the white dwarf and thus we have not included it in our figure.} Via the $S$-Higgs mixing, the hot environment in the interior of stars can abundantly create $S$ particles~\cite{Krnjaic:2015mbs,Dev:2020eam,Dev:2020jkh} (similar to axions and ALPs~\cite{Dicus1978,Raffelt:1990yz}), and the luminosity of the star would therefore be affected. Note that the stellar limit is reported only for $m_S > 100$ eV in Ref.~\cite{Dev:2020jkh}, it should also be able to constrain $m_S < 100$ eV to some extent and overlaps with our Leo T limit. There are also independent limits from precision measurements of Higgs and mesons that exclude $\theta \gtrsim 10^{-4}$~\cite{CMS:2018yfx,Dev:2020jkh} which are however relatively weaker than the limits in Fig.~\ref{fig:scalarLimits}.

The limit from Leo~T exceeds other limits in the 30$-$100 eV mass range, however stellar limits, if extrapolated to sub-keV masses, might be stronger. Our analysis can be easily translated to other scalar DM that decays to two photons. For instance, the dark glueball DM can decay to two photons with the decay rate controlled by parameters such as the dark confinement scale and the number of dark colors~\cite{Soni:2017nlm}. We anticipate the temperature of Leo~T would produce strong bounds for sub-keV dark glueballs as well.

%------------------------------------------------------------------------
\subsubsection{Sterile neutrinos}
\label{sec:nu}

Sterile neutrinos, especially those with keV-scale masses, are widely studied as a DM candidate and simultaneously, as a possible explanation of the unidentified X-ray lines at 3.5 keV (for detailed reviews, see e.g.~\cite{Merle:2013gea,Boyarsky:2018tvu}). Models of sterile neutrinos are normally parametrized by the mass of the sterile neutrino, $m_\chi$, and the mixing angle $\theta$ between sterile and ordinary neutrinos. In general, $\theta$ cannot be too large or the Universe would be overclosed by sterile neutrinos, as far as the known production mechanisms such as non-resonant~\cite{Dodelson:1993je} and resonant production~\cite{Shi:1998km} are concerned. Refs.~\cite{Gelmini:2019wfp,Gelmini:2019clw}, however, point out that these limits rely on the assumption of a standard cosmology and thus can be evaded with a non-standard cosmology prior to BBN.

We are therefore motivated to study robust observational limits on sterile neutrinos. The main model independent limits on $\theta$ come from the (non)observation of photons produced from their decays. At one-loop level, a sterile neutrino $\chi$ can decay to an ordinary neutrino $\nu$ and a photon, and the decay rate is~\cite{Boyarsky:2018tvu}
\begin{equation}
    \Gamma_{\chi\to \nu\gamma} = 5.5\times 10^{-22}\theta^2 \qty(\frac{m_\chi}{1\,\mathrm{keV}})^5\,\mathrm{s}^{-1},
\end{equation}
where $\theta^2=|\theta_e|^2+|\theta_\mu|^2+|\theta_\tau|^2$ is the total mixing to all the three ordinary flavors.

\begin{figure}[tb]
\centering
\includegraphics[width=0.45\textwidth,keepaspectratio=true]{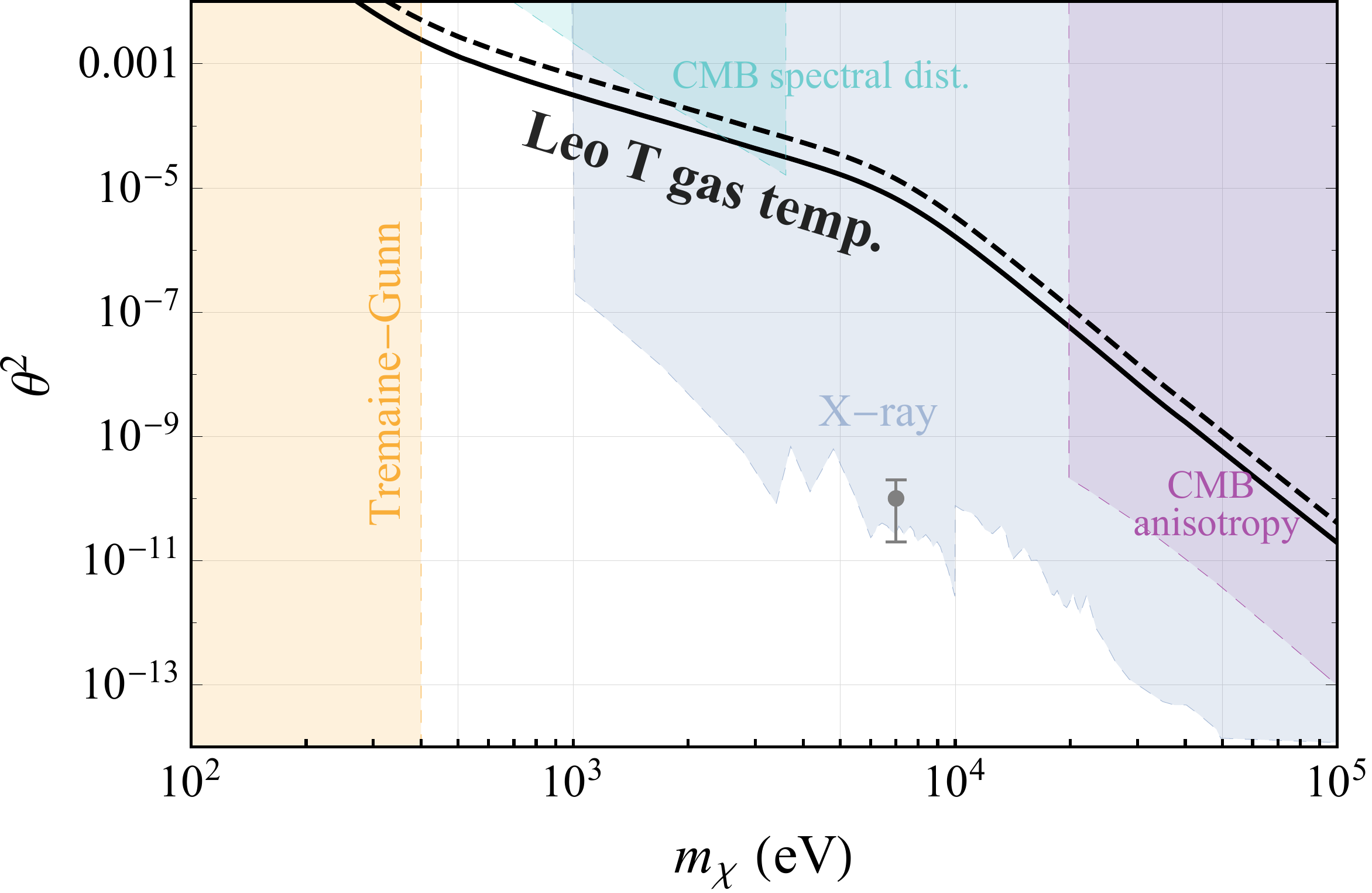}
\caption{Upper limits on the sterile neutrino mixing parameter $\theta^2$ as a function of $m_\chi$. We show limits from Leo~T in black, Tremaine-Gunn bound~\cite{TGbound} in orange, X-ray searches~\cite{Ess13,Gew17,Boyarsky:2018tvu,Foster:2021ngm} in gray, CMB spectral distortions~\cite{BolChl20} in cyan, and CMB anisotropy~\cite{Sla16} in violet. The dashed line shows the Leo T limit using the more conservative gas temperature estimate $T=$7552~K. The data point with error bars shows the sterile neutrino explanation of the 3.5 keV X-ray line excess~\cite{Boyarsky:2018tvu}.}
\label{fig:nuLimits}
\end{figure}

We again directly use all the limits presented in Fig.~\ref{fig:ModelIndep} to constrain $\theta$ in Fig.~\ref{fig:nuLimits}, except that $\taumin$ needs to be halved as only one photon is produced from the decay. We do not show the bound from Ref.~\cite{Nad21}, which is derived specifically for the case of resonantly produced sterile neutrinos. Aside from the photon-related limits, there is a universal lower limit on the mass of fermionic DM called the Tremaine-Gunn bound~\cite{TGbound}. This bound rules out the possibility of fermions lighter than 400 eV making up the entirety of DM based on dwarf galaxy phase space arguments.

The Leo~T limit covers the weakly constrained parameter space between X-ray searches and the Tremaine-Gunn bound, from 400 eV to 1 keV. Incidentally, this mass range coincides with the boundary between warm DM and cold DM. Warm DM has relatively long free-streaming length which would suppress structure formation, and its feasibility has been extensively studied using cosmological simulations. While the presence of small-scale structures disfavors a low-mass warm DM particle around keV~\cite{2012MNRAS.424..684S,Nad19,Nad21}, there has been no consensus conclusion about the precise value of the lowest allowed mass. In this respect, gas-rich dwarf galaxies can provide an independent venue to test warm sterile neutrino DM in parallel with cosmological methods.

%------------------------------------------------------------------------
\subsubsection{Dark baryons}

The next model we consider is a dark neutron $\chi$ that mixes with the ordinary neutron $n$. At the level of hadrons, the effective Lagrangian for the mixing can be written by $\lagr \supset \delta (\bar{\chi} n + \bar{n}\chi)$, where the parameter $\delta$ is determined by the UV completion of the model~\cite{McKeen:2015cuz}. If $|\delta| \ll |m_n-m_\chi|$, by diagonalizing the mass matrix we can see that the mixing induces the following neutron-dark neutron transition~\cite{McKeen:2020oyr}
\begin{equation}
    \lagr \supset \frac{\mu_n}{2} \theta \bar{\chi} \sigma_{\mu\nu}F^{\mu\mu} n + \mathrm{c.c.},
\end{equation}
where $\mu_n$ is the magnetic dipole moment of neutrons, and $\theta = \delta/|m_n-m_\chi|$. The dark neutron model has received a large attention as it provides a simple setting to address the neutron lifetime anomaly~\cite{Fornal:2018eol,Cline:2018ami}, baryogenesis~\cite{McKeen:2015cuz,Aitken:2017wie} and the XENON1T anomaly~\cite{McKeen:2020vpf}, with potential connections to DM. 

In this study, we consider the case of heavy dark neutron DM, i.e., $m_\chi > m_n$. The two decay modes are $\chi \to n\gamma$ and $pe\nu$, with the former dominating because it does not require weak boson mediation. The decay rate of $\chi\to n\gamma$ is given by~\cite{McKeen:2020oyr}
\begin{equation}
    \Gamma_{\chi\to n\gamma}= 4.5\times 10^{16}\theta^2 \left(\frac{E_\gamma}{10\,\mathrm{MeV}}\right)^3\, \mathrm{s}^{-1},
\end{equation}
with $E_\gamma$ being the energy of the photon $E_\gamma=m_\chi(1-m_n^2/m_\chi^2)/2 \simeq m_\chi - m_n$. 

Let us first derive the limit on $\theta$ from gas heating of Leo~T. As long as the dark neutron is not tightly degenerate with the ordinary neutron, the typical energy of the photon from dark neutron decay is above MeV. The corresponding opacity factor $1-e^{-\lambda_\gamma}$ in Fig.~\ref{fig:opacity} is therefore exceedingly small. As a conservative estimation, we neglect the heating due to photons. The main heating source is the neutrons from dark neutron decay. These neutrons would most likely  $\beta$-decay into electrons before they collide with gas particles because the neutron decay rate $\Gamma_n$ is much larger than the scattering rate $n_{\mathrm{gas}} \expval{\sigma v} $ for neutron-nucleus cross sections. The electrons would then interact with the gas and deposit heat as we have discussed in Sec.~\ref{sec:eEnergy}.

\begin{figure}[tb]
\centering
\includegraphics[width=0.45\textwidth,keepaspectratio=true]{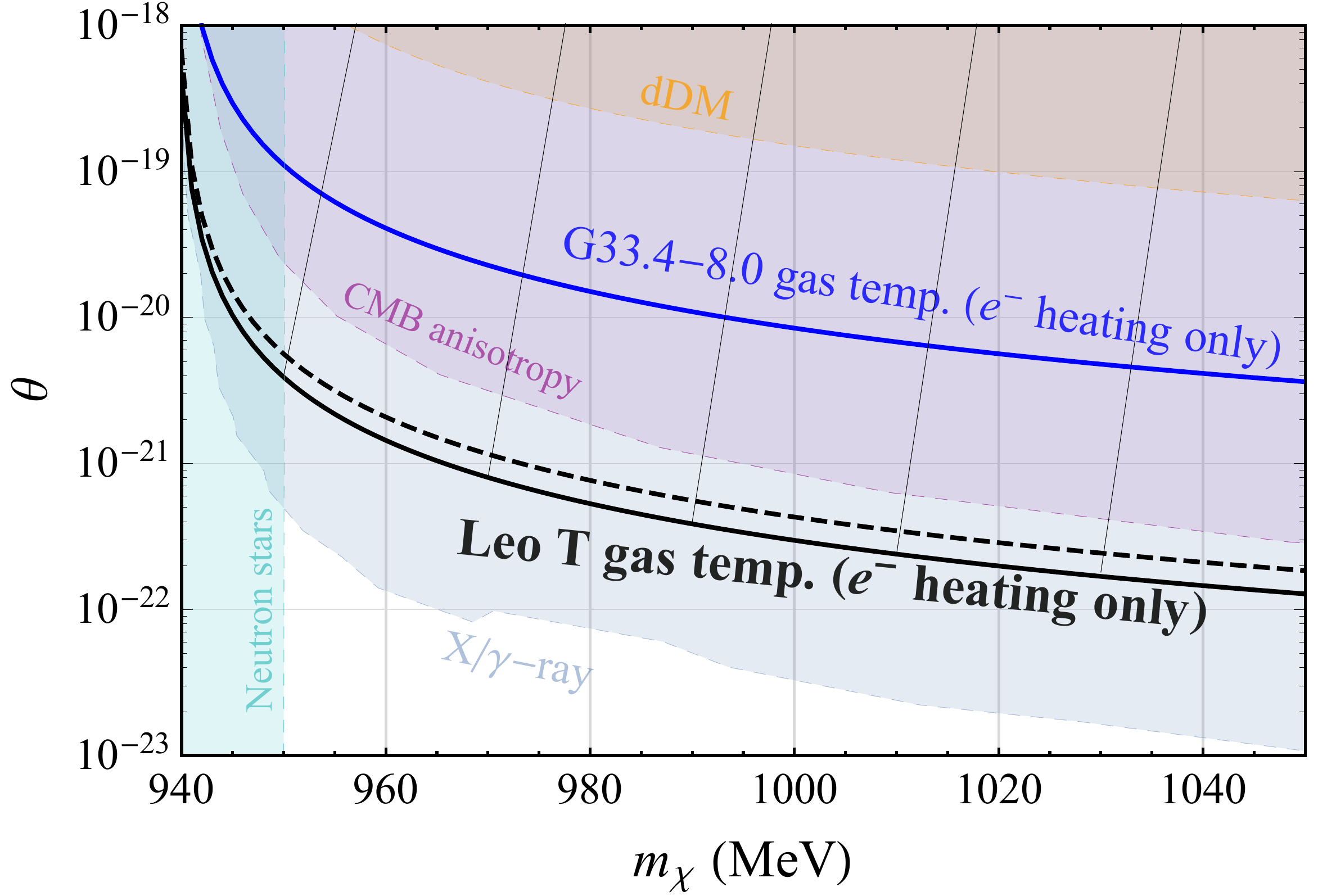}
\caption{Upper limits on the neutron-dark neutron mixing angle $\theta$ as a function of $m_\chi$. The black and blue curves are obtained from the temperature of Leo~T and MW gas clouds. The Leo~T limit is conservative as we neglect the heating by photons. The dashed line shows the Leo T limit using the more conservative gas temperature estimate $T=$7552~K. Also shown in the plot are general dDM constraints~\cite{Audren:2014bca,Poulin:2016nat} in orange, X/$\gamma$-ray searches~\cite{Ess13,Gew17,Boyarsky:2018tvu,Foster:2021ngm} in gray, CMB anisotropy~\cite{Sla16} in violet, and the temperature of cold neutron stars~\cite{McKeen:2020oyr,McKeen:2021jbh} in cyan. The X/$\gamma$-ray and CMB anisotropy limits are recast by Eq.~\ref{eqn:recast}.}
\label{fig:DarkNeutron}
\end{figure}

A small complication here is caused by the fact that the energy of these electrons is not a constant. Fortunately, since the neutron has very low kinetic energy, the spectrum of electrons from the $\beta$-decay can be studied in the rest frame of neutrons. To leading order, the differential decay width of neutrons is
\begin{equation}
    \dv{\Gamma_n}{E_e}=\frac{1+3\lambda^2}{2\pi^3}|V_{ud}|^2 G_F^2 \sqrt{E_e^2-m_e^2}E_e(m_n-m_p-E_e)^2,
\end{equation}
where $\lambda=-1.27$ is the vector-axial ratio, and $V_{ud}=0.97$ is the CKM matrix element. The probability density of electrons with energy $E_e$ is thus given by $\beta(E_e)\equiv\Gamma_n^{-1} d\Gamma_n/dE_e$. The volume averaged heating rate is then
\begin{equation}
    \dot{Q} = \frac{1}{r^3_\mathrm{WNM}/3}\int  drdE_e\,r^2 n_{\chi} E_{\mathrm{heat}} \beta(E_e) \Gamma_{\chi},
\end{equation}
where $E_{\mathrm{heat}} = f_e(E_e - m_e) (E_e - m_e) (1-e^{-\lambda_e})$, and the integration on $E_e$ ranges from $m_e$ to $m_n-m_p$. By comparison with the cooling rate of Leo~T, we obtain the limit from Leo~T on $\theta$ in Fig.~\ref{fig:DarkNeutron}. The similar limit from MW gas clouds is depicted by the blue curve.

In Fig.~\ref{fig:DarkNeutron} we show additional limits from dDM~\cite{Audren:2014bca,Poulin:2016nat} in orange, X/$\gamma$-ray searches~\cite{Ess13} in gray, CMB anisotropy~\cite{Sla16} in violet, and the temperature of cold neutron stars~\cite{McKeen:2020oyr,McKeen:2021jbh} in cyan. The X/$\gamma$-ray and CMB limits are obtained as follows. The original limits in Refs.~\cite{Ess13,Sla16} (also in Fig.~\ref{fig:ModelIndep}) are limits on the lifetime of DM decaying to two photons, and would not directly apply to dark neutrons. Given a set of limits $(m, \taumin_{\gamma\gamma})$, which prescribes the minimal lifetime of DM decaying to $\gamma\gamma$ for DM mass $m$, we can map it to $(m_\chi,\taumin_{\chi\to n\gamma})$ by
\begin{equation}
\label{eqn:recast}
    (m_\chi,\taumin_{\chi\to n\gamma}) = \left(\frac{m}{2}+m_n, \frac{m}{2m_\chi}\taumin_{\gamma\gamma}\right).
\end{equation}
In the $\gamma\gamma$ decay mode, each photon has energy $E_\gamma = m/2$ and therefore dark neutrons with mass $m_\chi \simeq m_n + m/2$ will produce photons with the same energy $E_\gamma$. The factor $m/(2m_\chi)$ accounts for the scaling in the number density of photons from $\chi$ decay relative to the $\gamma \gamma$ decay case.

The gray and violet region in Fig.~\ref{fig:DarkNeutron} show the excluded values of $\theta$ from X/$\gamma$-ray searches and CMB anisotropy reinterpreted via Eqn.~\ref{eqn:recast}. Due to the dilution effect of $m/(2m_\chi)$ in Eqn.~\ref{eqn:recast}, our CMB limit on $\theta$ is weaker than the one reported in Ref.~\cite{McKeen:2020oyr}. We also direct readers to Ref.~\cite{McKeen:2020oyr} where limits on dark neutron DM with other primordial densities and BBN limits on non-DM dark neutrons are derived. In general, although the Leo~T limit on dark neutrons is less constraining than X/$\gamma$-ray limits, they rely on different products from dark neutron decay and are therefore complementary. 

% In order to recast these limits and constrain the decay of $\chi$, given a set of model independent limits on the mass and decay lifetime $(m_\DM,\taumin_{\DM\to\gamma\gamma})$, we need to map the relation between $m_\DM$ and $m_\chi$ such that they produce the same photon energy flux \DW{The notation $m_{DM}$ in this case is a bit confusing, can you change it?}. The translated constraint $(m_\chi,\taumin_{\chi\to n\gamma})$ can be found by
% \begin{equation}
% \label{eqn:recast}
%     (m_\chi,\taumin_{\chi\to n\gamma}) = \left(\frac{m_\DM}{2}+m_n, \frac{m_\DM}{2m_\chi}\taumin_{\DM\to\gamma\gamma}\right).
% \end{equation}
% The light gray and violet region in Fig.~\ref{fig:DarkNeutron} shows the excluded values of $\theta$ from X/$\gamma$-ray searches and CMB anisotropy reinterpreted via Eqn.~\ref{eqn:recast}. Due to the dilution effect in Eqn.~\ref{eqn:recast}, our CMB anisotropy limit on $\theta$ is weaker than the one reported in Ref.~\cite{McKeen:2020oyr}. The gas cloud limits on dark neutrons are generally less constraining than X/$\gamma$-ray limits, but provide complementary probes to the model. 

We also note that there are models of DM particles carrying two units of baryon number~\cite{Farrar:2018hac,Farrar:2020zeo, Heeck:2020nbq}. If such particles can decay to two neutrons, our methodology can be tailored to constrain them. We will discuss this scenario in a future work~\cite{FW21}.

%------------------------------------------------------------------------
\subsubsection{Excited DM states and other models}
%In the earlier section we focused our discussion on the decay of a particle into two photons. Another possibility is the de-excitation of a meta-stable excited state of DM associated with the emission on one photon. Models of excited states in DM have been considered in connection with the 511 keV line observed by INTEGRAL and the DAMA anomaly~\cite{Fin07,Fin09}. Ref.~\cite{Bar20} uses to explain the XENON1T anomaly.

Models of excited or inelastic DM assume the DM particle $\chi$ has an excited state $\chi^*$~\cite{Fin07,Fin09}. The scenario enables a rich phenomenology to explain a few anomalies observed by direct detection experiments, including DAMA and XENON1T~\cite{Fin07,Fin09,Bar20}. Furthermore, depending on the magnitude of the the mass splitting $\delta = m_{\chi^*}-m_\chi$, the excited state $\chi^*$ can de-excite via transitions $\chi^* \to \chi e^+ e^-$ or $\chi^* \to \chi \gamma$, providing potential solutions to the Integral 511 keV and 3.5 keV X-ray excess~\cite{Fin07,fin14}.

The de-excitaton processes $\chi^* \to \chi e^+ e^-$ and $\chi^* \to \chi \gamma$ can be constrained by Leo~T. For example, the constraint on the lifetime of $\chi^* \to \chi \gamma$ is identical to sterile neutrinos in Sec.~\ref{sec:nu}. The decay rate of $\chi^* \to \chi \gamma$ in general depends on the mass splitting $\delta$, the fraction of DM that is in the excited state, as well as the model dependent details of how the transition is generated (i.e. the coupling constants). For benchmark models with $\delta \lesssim 1$ keV, we expect Leo~T to impose strong limits.

%------------------------------------------------------------------------

\section{Discussion}
\label{sec:Discussion}

Let us now discuss the cases in which our DM limits are complementary to existing astrophysical limits in the literature.
Decays and annihilations of DM have already been probed by searching for resulting optical, X-ray or $\gamma$-ray photons from astrophysical systems. However, photons in the UV and soft X-ray ranges are extremely difficult to probe in direct searches as these are easily scattered/absorbed in a typical astrophysical medium. This is precisely the range where Leo T gives strong constraints (see Fig.~\ref{fig:opacity}). Apart from astrophysical searches, early-Universe probes of CMB anisotropies and CMB spectral distortions have been used to constrain DM annihilation and decay \cite{Fin12,Sla16,Liu:2016cnk,HaiChlKam15,BolChl20,Ali21} (as they are also sensitive probes of non-standard heating/ionization of gas). 
Our constraints are complementary to them because the systematics and assumptions in our analysis are entirely different (e.g., we have no assumptions about cosmology in our analysis).

It is worth discussing the observational criteria that make Leo T a strong calorimetric probe of DM. We believe the following criteria can also be used to identify other astrophysical systems which could give comparable or stronger DM limits:\\
$(i)$ the DM density surrounding the \HI gas should be large.\\ $(ii)$ The gas cooling rate of the system should be low (which implies low values of \HI number density, gas temperature, and gas metal fraction).\\ $(iii)$ The gas should be in a relatively steady state (the system should not be undergoing a merger or significant accretion).

Let us now discuss the susceptibility of our DM bounds to the uncertainties in modeling the DM halo of Leo T. As discussed in Sec.~\ref{sec:LeoT}, we adopted the best-fit model by Ref.~\cite{faerman13} for the DM halo. Ref.~\cite{faerman13} also reports a range of fitted DM halo parameters with 3$\sigma$ errors. We find that the parameters within this 3$\sigma$ range, which give the worst possible weakening, change our DM decay lifetime bounds only by $\lesssim$5\% (DM annihilation cross section bounds change by $\lesssim$10\%). There is one additional point worth discussing. Given the \HI line of sight velocity data of Leo T, there is no measurable rotational motion of \HI gas \cite{adams17}. However, it could be that the axis of rotation of the gas aligns with our line of sight. In this case, the model of Ref.~\cite{faerman13}, which used hydrostatic equilibrium to calculate the DM halo mass, needs to be revised, and the dynamical halo mass could increase significantly (which would make our DM bounds stronger). We also note that the model of Ref. \cite{faerman13} does not
take into account astrophysical heating and radiative cooling of the \HI gas. In a future study, we plan to perform hydrodynamic simulations of gas-rich dwarfs like Leo T which include thermal feedback from DM alongside the standard astrophysical heating and cooling effects.

\section{Conclusions}
\label{sec:Conclusions}
Some of the low-mass gas-rich dwarf galaxies located outside the virial radius of their host are relatively pristine systems and have ultra-low gas cooling rates. This makes the gas in the systems very sensitive to heat injection by annihilation or decay of dark matter (DM). We required that DM heat injection rate to be lower than the gas radiative cooling rate in the Leo~T dwarf galaxy and our primary results are as follows:\\ $(i)$ we set bounds on the decay lifetime of DM to $e^\pm$ (photons) which are stronger than all the previous literature
for $m_\textup{DM}\sim$ 1$-$10 MeV ($m_\textup{DM}\sim0.02-$1 keV) [see top panel of Fig.~\ref{fig:ModelIndep}].\\ $(ii)$ we constrain annihilation cross section of DM to $e^{\pm}$ comparable to constraints from CMB and X/$\gamma$-ray surveys (see bottom panel of Fig.~\ref{fig:ModelIndep}).\\ $(iii)$ we translate our constraints to specific DM models like axion-like particles (ALPs), Higgs portal scalars, sterile neutrinos and dark baryons (see Figs.~\ref{fig:Axion}, \ref{fig:scalarLimits}, \ref{fig:nuLimits} and \ref{fig:DarkNeutron} respectively) and obtain strong bounds.

% To conclude this section, we have considered a few particular decaying DM models, and obtained constraints on the parameter space based on the temperature of gas clouds in Leo~T. By comparison with other limits like CMB and X-ray searches, we discover that Leo~T limits set the strongest constraints in the sub-keV regime.

%\noindent\textit{Future work:} 
We plan to report limits from Leo~T on gas heating due to electrophilic axions and compact objects like primordial magnetic black holes in an upcoming work \cite{WadWaninprep20}. 
%Mergers of exotic compact objects can also inject energy in the gas-rich dwarf galaxies (e.g.~\cite{Hertzberg:2020dbk,Amin:2020vja,Kritos:2021nsf}) and limits from them will also be reported.
%For instance, electromagnetic signals from compact object mergers~\cite{Hertzberg:2020dbk,Amin:2020vja,Kritos:2021nsf} can also inject energy in the gas-rich dwarf galaxies.
% In an upcoming work~\cite{WadWaninprep20}, we will discuss other scenarios of energy injection from dark sectors.
Some other low-mass gas-rich dwarfs have properties similar to Leo~T and are also well-studied (e.g., Leo~P \cite{Ber15}); we plan to derive DM limits from them in a future work.
We note that some dwarfs recently discovered in Ref.~\cite{Jan19} could have even smaller gas cooling rates than Leo T. However, such an analysis requires follow-up measurements of their dynamical masses and gas kinematics, which are currently lacking.

%Although we have only considered decay to photons and e$^\pm$ in this paper, one could also constrain decays/annihilations to $\mu, \tau, b \bar{b}, etc$ and these might be considered in a future study.
Upcoming optical and 21cm surveys will find and characterize a much larger number of ultra-faint and gas-rich dwarfs than present, e.g., the Rubin observatory could detect dwarfs like Leo T ($M_V=-8$) up to $\sim5$ Mpc \cite{Mut21}. This might potentially result in detection of hundreds of galaxies similar to Leo T and can enable even more stringent probes of heat exchange due to DM.

The mathematica code associated with this paper and the data files for the plots are publicly available online \href{https://github.com/JayWadekar/Gas_rich_dwarfs}{\faGithub}.\footnote{\href{https://github.com/JayWadekar/Gas_rich_dwarfs}{\textcolor{blue}{https://github.com/JayWadekar/Gas\_rich\_dwarfs}}.}

% For instance, electromagnetic signals from compact object mergers~\cite{Hertzberg:2020dbk,Amin:2020vja,Kritos:2021nsf} can also inject energy in the gas-rich dwarf galaxies.

%Astrophysical systems with low radiative cooling rates are sensitive to energy transfer by DM collisions and should be further explored to search for possible signs of DM interactions.
\acknowledgments

We thank Siyao Xu, Ken Van Tilburg, Jim Stone, Glennys Farrar, Hongwan Liu, Marco Muzio, Chris McKee, Cristina Mondino, Amiel Sternberg, Andrey Kravtsov and Mariangela Lisanti for useful discussions. We also thank Glennys Farrar, Hyungjin Kim, Cristina Mondino, Marco Muzio and Nirmal Raj for their comments on the draft of this paper. We thank the anonymous referee for their critical comments and various useful suggestions which improved the manuscript.
We thank Yakov Faerman and Shmuel Bialy for providing us the data for the best-fit model of Leo~T from Ref.~\cite{faerman13}.
We are also grateful to Hongwan Liu for collaboration on the initial stages of the paper.
DW gratefully acknowledges support from the Friends of the Institute for Advanced Study Membership. We have taken the data for exisiting limits for the axion-photon coupling from the public repository by Ciaran O'Hare: \url{https://github.com/cajohare/AxionLimits}

\appendix

\section{Details on HI gas cooling}
\label{apx:cooling}
In section~\ref{sec:Cooling}, we had presented an overview of the cooling of \HI gas in Leo~T.
In this Appendix, we present additional details on the calculation of the radiative cooling rate. For $T_\mathrm{gas}\sim 6000$ K, the collisional cooling is dominated by fine structure lines of metals in the gas and the rate is primarily dependent on the number density of metals, neutral hydrogen and electrons in the gas \cite{Maio07}. To model the electron density, one needs to take into account ionization of gas due to the UV/X-ray metagalactic background (e.g., \citep{hm12}) (the additional ionization enhances the metal line de-excitations, thereby increasing the gas cooling rate). In the case of high density gas, however, the gas can self-shield itself from the UV metagalactic radiation, which results in lowering the cooling rate. An ideal way of accounting for the metagalactic ionization and the self-shielding effects in an astrophysical system is to model the system using a radiative transfer code.

\WF\ calculated the cooling rate using Eq.~\ref{eq:grackle} and obtained the cooling function $\Lambda(T)$ from the widely-used astrophysical radiative cooling software Grackle \cite{grackle}. Grackle however does not do a full radiative transfer computation but uses the following approximations. It uses pre-computed cooling tables from the \textsc{Cloudy} \cite{cloudy} code for a gas irradiated with the metagalactic background. However, the \textsc{Cloudy} tables were calculated under the plane-parallel assumption (i.e for a plane-parallel gas slab illuminated by a perpendicular beam), and are therefore an approximation for the case of Leo~T which has a nearly spherical geometry. One other assumption in Eq.~\ref{eq:grackle} is that the cooling function $\Lambda$ depends only on the temperature of the gas and not on its ionization state or the number density.
Furthermore, Grackle also includes an approximate prescription for the self-shielding effect using fitting functions from \cite{Rahmati13}. 
%Note however that the aforementioned assumptions are not supposed to have minor 
%Therefore, the results derived from using the Grackle cooling function involve the above assumptions.

Instead of using the aforementioned assumptions, one can obtain a more accurate estimate of the cooling of the gas by
using a radiative transfer code and then solving for the metal cooling transitions. Ref.~\cite{faerman13} performed a full radiative transfer computation to model the ionization profile of Leo~T (as shown in Fig.~\ref{fig:LeoTobs}). \Kim\ used this model and explicitly solved the radiative cooling transitions of the most important metals to obtain their estimate: $\dot{C}\simeq 7 \times 10^{-30}$ GeV cm$^{-3}$ s$^{-1}$, which we used in our study. Note that this is slightly higher (factor of $\sim 1.8$) than the one obtained in \WF\ using the approximate cooling estimate from Eq.~\ref{eq:grackle} and the Grackle cooling function.

Apart from metal line cooling, there is also the cooling due to emission of Lyman-$\alpha$ photons when collisions between H atoms can excite the electrons. This channel is however exponentially suppressed when the gas temperature is lower than that corresponding to the difference in the first two energy levels of the H atom (i.e., 10.2 eV); the corresponding cooling rate is given by \cite{Dal72,Kim20}
%As $T_\mathrm{gas}$ increases, 
%The Lyman$\alpha$ cooling rate is given by 
\be
\frac{\dot{\mathcal{C}}_{\mathrm{Ly} \alpha}}{\mathrm{erg/cm}^{3}/\mathrm{s}}=n_\mathrm{HI}\, n_e\left[ 7.5 \times 10^{-19} \frac{e^{-118348\, \mathrm{K}/T}}{1+\left(\frac{T}{10^5\mathrm{K}}\right)^{1/2}} \right] 
\ee
Given the abundance of H and metals in the WNM of Leo T, we find that the Ly-$\alpha$ cooling starts dominating over metal line cooling around $\sim 8000$ K, but is negligible for the observed gas temperature: $T<6100$ K.
As discussed in section~\ref{sec:GasTemp}, we calculate limits for the more conservative case of temperature of the gas (i.e., $T_\mathrm{gas}=7552$ K). The corresponding $\dot{C}$ becomes $\simeq 1.46 \times 10^{-29}$ GeV cm$^{-3}$ s$^{-1}$ upon accounting for both the metal line and Ly-$\alpha$ cooling channels (with both contributing roughly equally).
%(which was $\sim 3.8 \times 10^{-30}$ erg cm$^{-3}$ s$^{-1}$).
%taking into account the ionization and density profiles

\subsection{Leo T gas metal abundances}

%\footnote{\href{https://grackle.readthedocs.io/}{\textcolor{blue}{https://grackle.readthedocs.io/}}} .
For calculating the metal cooling rate, we have assumed that the gas metal abundance relative to solar is given by the value derived using stellar spectroscopy ($10^\mathrm{[Fe/H]}=10^{-1.74}$) \cite{Kir13}. In this sub-section, we discuss some of the assumptions involved in this regard. A more appropriate metallicity value to use in the gas cooling calculation is the one calculated directly from the gas instead of stellar spectroscopy. However, the gas-phase metallicity has not been directly measured in Leo T because of lack of an extended HII region in the gas (needed to get spectra). It is therefore worth comparing the the metallicity value of Leo T with the prediction from metallicity scaling relations and also relative to other dwarf galaxies with similar properties.
Using the mean luminosity-metallicity relation from \cite{Ski13_LeoP} or stellar mass-metallicity relations from \cite{Ber12}, we find the metal fraction to be $\sim 10^{-2}$ for Leo T (we use the following measured values for Leo T: $M_V=-8$ and $M_*\sim 10^5\, M_\odot$ \cite{Zou21}).

Leo P is another gas-rich low-mass dwarf galaxy with properties broadly similar to Leo T. For Leo P, however, a HII region has been detected and the gas metal abundance derived from oxygen lines is $10^{-1.75}$ \cite{Ski13_LeoP}. It is also worth mentioning that the metal abundance derived from stellar spectroscopy in Leo P ($10^\mathrm{[Fe/H]}=10^{-1.8}$) is fairly well consistent with the gas metal abundance.
The stellar mass of Leo T is smaller than Leo P (by a factor of $\sim 5$), we expect the Leo T metal abundance to be lower than Leo P. Hence, the value of Leo T metallicity used in this paper ($10^{-1.74}$) is conservative.
%\new{It is worth mentioning that metals can also be removed from \HI gas due to capture by dust particles (also known as the interstellar depletion effect).}

It is worth noting that the cooling rate calculation by \Kim\ assumes that the ratio of chemical abundances of various $\alpha$-elements (e.g., oxygen or carbon) to the iron peak elements [$\alpha$/Fe] is $\sim 0$ in Leo T. As we argued earlier, $\alpha$-abundances have not been measured for Leo T but for Leo P, the oxygen abundance is indeed similar to Fe abundance. See also section 5.1 of \Kim\ where they provide an alternative justification for [$\alpha$/Fe] being $\sim 0$ for Leo T.

In a general scenario, gas metallicity can be lower than stellar metallicity due to capture of metals in gas by dust particles (i.e., the interstellar depletion effect). 
This effect is however negligible for very low-metallicity gas in galaxies like Leo~T (where the dust fraction is expected to be low) \cite{Bia19,Kim20}.

% Note that stars inside Leo~T can emit ionizing radiation and increase the local electron density, and thereby the radiative cooling rate. However, the stellar radiation also heats the gas and therefore the overall effect of local stellar radiation on our limits based on heating due to DM is expected to not be significant.

\section{Details on $e^{\pm}$ confinement}

\subsection{\e streaming velocity}
\label{StreamingVel}
In section~\ref{sec:eEnergy}, we had assumed that the relativistic \e with energies MeV$-$GeV produce \alf waves, which in turn scatter them, and therefore the \e move with \alf velocity ($v_A$) which is much smaller than $c$. 
In this sub-section, we explicitly calculate the streaming velocity ($v_\textup{stream}$) of \e and show that it is indeed close to $v_A$.
$v_\textup{stream}$ can be obtained by balancing
the growth rate of \alf waves due to relativistically streaming $e^\pm$ with the wave damping rate due to ion-neutral collisions.
%The primary mechanism of decay of \alf waves is
The ion-neutral damping rate is given by (see Eq.~(A3) of \cite{Fer88}):
\be
\Gamma_\mathrm{damping} \simeq 10^{-10} \mathrm{s}^{-1} (1-0.9\, x_\mathrm{p}) \bigg( \frac{n_\textup{HI}}{0.06\, \mathrm{cm}^{-3}} \bigg)
\ee
for $x_\mathrm{p}\gtrsim 10^{-2}$, where $x_\mathrm{p}$ is the ionization fraction. In the frame of reference moving with the Alfven waves, the wave growth rate is \cite{Kul69,Ces80}:
\be
\Gamma_\mathrm{growth}\simeq \frac{c}{r^0_g} \frac{n(r_g\geq r^0_g)}{n^*} \left(\frac{v_\mathrm{stream}}{v_A}-1\right)
\ee
where $n(r_g\geq r^0_g)$ is the number of charged particles having gyration radius larger than $r^0_g=\gamma m_e c/eB$, $n^*$ is the number density of charged particles, $v_\mathrm{stream}$ is the streaming velocity of $e^{\pm}$ along the magnetic field lines. $n(r_g\geq r^0_g)$ has two contributions: the \e emitted by DM, and the mildly to fully relativistic cosmic rays produced due to star formation and supernova activity (see e.g.,~\cite{Wen74}). Performing a rough estimate for the first source: $n(r_g\geq r^0_g)\sim2\, n_\DM$ \tc$/\tau_\DM $  (we use $\tau_\DM$ from the bound in Fig.~\ref{fig:ModelIndep}), we already find $(v_\mathrm{stream}/ v_A -1)\ll 1$ and therefore $v_\mathrm{stream}\simeq v_A$ for the range of values covered in our plots.

It is also worth mentioning that the \alf waves travel along the anisotropy direction of the \e distribution. As the DM density falls radially, the gradient of the generated \e distribution is also along the radial direction. Furthermore, the star formation profile in Leo T is also concentrated towards its center, so the \e produced due to it will also have a radial gradient. 

% However, when the wave growth due to the
% streaming instability is inhibited by some damping process, such as turbulent damping, the coupling of CRs to the
% gas is weaker and their effective propagation speed faster than the \alf speed.

% \be
% \frac{v_\mathrm{stream}}{v_A} - 1 = 2.3 \left(\frac{x_e}{10^{-2}}\right) \bigg( \frac{n_\textup{HI}}{0.06\, \mathrm{cm}^{-3}} \bigg) \mathcal{M}_A
% \ee

%Because the DM density profile falls with the distance from the center, the generated e$^\pm$ have a net radial bulk radial velocity outward from the center.

\subsection{Diffusion of magnetic field lines}
\label{FieldDiffusion}
In Sec.~\ref{sec:eEnergy}, we argued that the field lines are tangled rather than straight, and are also constantly being re-arranged by the turbulence. Thus even though \e stream along field lines, the motion of \e resembles diffusion as the field lines themselves are in a random walk configuration. In this Appendix, we quantify this diffusive motion.

There is no rotation observed in Leo~T so the field orientation is therefore expected to be random rather than ordered.
%$e^{\pm}$ will therefore undergo a diffusive rather than a coherent motion. 
The coherence length of the field is given by  $L_\mathrm{coh}\sim L_\textup{turb}\, \textup{min}(1,\mathcal{M}_A^{-3})$ \cite{Bru07}, where $L_\textup{turb}$ is the scale at which ISM is stirred due to supernova explosions or star formation (i.e. the driving scale of the turbulence). $\mathcal{M}_A$ is the \alf Mach number of the turbulence at its injection scale and we discuss it in Sec.~\ref{apx:MA} below. $L_\textup{turb}$ is typically 50$-$100pc in spiral galaxies, however it can be much smaller (i.e. 1$-$20 pc) for dwarf galaxies \cite{Min96}. We conservatively use the value 100 pc in our calculations. The diffusion coefficient therefore becomes $D\simeq  v_A L_\mathrm{coh}$/3 for the $\mathcal{M}_A\geq 1$ case \cite{Kru20}
%\footnote{A more general definition is $D\simeq  v_\mathrm{stream} L_\mathrm{coh}$/3 and is usually energy dependent at high energy use is not applicable in the $E_e\gg$ GeV regime as the energy density of \e available to excite magnetic field fluctuations at a given gyroradius scale declines sufficiently and $v_\mathrm{stream}$ can exceed $v_A$, therefore the diffusion coefficient also becomes energy dependent.}
, which we use to calculate the confinement timescale in the WNM as

\be\begin{split}
t_\mathrm{confine} &\simeq \frac{r^2_\mathrm{WNM}}{v_A L_\mathrm{coh}}\simeq 19\, \frac{1\, \mu \mathrm{G}}{B} \frac{100\, \mathrm{pc}}{L_\mathrm{turb}} \left(\frac{\mathcal{M}_A}{1}\right)^3  \mathrm{Myr}
\end{split}\label{tconfine}\ee

It is worth mentioning that our estimate of the confinement time is similar to that of \Kim, who report $\sim 9$ Myr (\Kim\ does not consider the effects of turbulence and therefore derives a slightly more conservative time estimate).
Note that we do not considering scattering of $e^\pm$ with magnetic field inhomogeneities (e.g.~\cite{Laz21,Xu21}) and including this effect will further increase $t_\mathrm{confine}$.

\subsubsection{Calculation of $\mathcal{M}_A$}
\label{apx:MA}
The value of $\mathcal{M}_A$ governs the degree to which the field lines are re-arranged by the turbulence. In the absence of a large-scale ordered magnetic field, as is the case for Leo T, $\mathcal{M}_A$ is expected to be larger than one. We adopt the conservative value of $\mathcal{M}_A=1$ in this paper.

Let us still discuss the explicit form of $\mathcal{M}_A$ in case robust estimates from observations of Leo~T or simulations are available in the future. $\mathcal{M}_A$ is given by \cite{Kru20}
\be\begin{split}
\mathcal{M}_A =& \frac{v_\mathrm{turb}}{B/\sqrt{4\pi \rho}} = 1 \frac{1\, \mu \mathrm{G}}{B} \left(\frac{\rho_\mathrm{total}}{0.075\, \mathrm{GeV\, cm}^{-3}}\right)^{\frac{1}{2}} \frac{v_\mathrm{turb}}{7.7\, \mathrm{km/s}}
\end{split}\ee
where $\rho$ is the total matter density (ions + neutrals), and $v_\mathrm{turb}$ be the turbulent velocity of Alfv$\acute{\text{e}}$nic modes at the injection scale of the turbulence.
The relation of $v_\textup{turb}$ to the 3D ISM velocity dispersion ($\sigma$) depends on how the turbulent energy is partitioned between Alfv$\acute{\text{e}}$nic, slow and fast modes \cite{Cho02}.  It is likely that there is somewhat larger energy in Alfv$\acute{\text{e}}$nic modes as compared to the other two modes and therefore $v_\textup{turb}> \sigma/\sqrt{3}$ (for reference, Ref.~\cite{Kru20} estimate likely values could be $v_\textup{turb}\sim \sigma/\sqrt{2}$, and $\mathcal{M}_A \sim 2$ based on inputs from simulations, see also \cite{Cro21}). 
%We have adopted the turbulent velocity of the Alfv$\acute{\text{e}}$nic modes to be $\sigma/\sqrt{2}$ at the injection scale assuming half the turbulent energy present in Alfv$\acute{\text{e}}$nic modes \cite{Kru20}. 

\section{Opacity estimates of the MW co-rotating cloud}
\label{apx:opacity}
For our estimates, we use the size of the G33.4$-$8.0 cloud to be the geometric mean of the two observed dimensions from \cite{PidLock15}: $\sqrt{24\times 8}$ = 13.9 pc.
We substitute this value into Eq.~\ref{eq:photoOpacity} to obtain the photon optical depth. Let us now calculate the opacity of the cloud to \e. We use the diffusion coefficient in the Milky Way for a particle with energy $E$ as $D\sim 3\times 10^{27} \left(\frac{E}{1\, \mathrm{GeV}}\right)^{1/2}$ cm$^2$ s$^{-1}$ from \cite{Kru20}.
This gives the confinement time as $t_\mathrm{confine} = r^2_\mathrm{cloud}/(6D)$, which we substitute in Eq.~\ref{eq:eOpacity} to get the opacity.
It is worth mentioning that annihilating/decaying dark matter can create a diffuse flux of photons or \e within the halo of the MW. Such diffuse flux can also heat the gas in the MW clouds in addition to heating from emission from DM within the cloud, but we have not considered this additional heating effect in this paper.

\bibliographystyle{apsrev4-1}
\bibliography{leoT}

\end{document}